\documentclass[prd,showpacs,nofootinbib,preprintnumbers]{revtex4}
\usepackage{epsfig,graphics,color}
\usepackage{subfigure}
\usepackage{bbm}
\usepackage{amssymb}
\usepackage{amsmath}
\usepackage{mathrsfs}
\usepackage{bbold}
\usepackage{rotating}
\usepackage{lscape}
\begin{document}
\title{Chiral symmetry breaking from two-loop effective potential of the holographic non-local NJL model} %[arXiv:hep-th/]}
\author{Piyabut Burikham$^{1,3}$}
\author{Daris Samart$^{2,3}$}
\author{Suppiya Siranan$^{2}$}

\affiliation{$^{1}$Department of Physics, Faculty of Science, Chulalongkorn University, Phyathai Road, Bangkok 10330, Thailand}
\affiliation{$^{2}$Department of Applied Physics, Faculty of Sciences and Liberal arts, Rajamangala University of Technology Isan, Nakhon Ratchasima, 30000, Thailand}
\affiliation{$^{3}$Thailand Center of Excellence in Physics (ThEP), Commission on Higher Education, Bangkok 10400, Thailand}

%\emailAdd{piyabut@gmail.com}
%\emailAdd{daris.sa@rmuti.ac.th}
%\emailAdd{suppiya.si@rmuti.ac.th}
%%%%%%%%%%%%%%%%%%%%%%%%%%%%%%%%%%%%%%%
\begin{abstract}
We calculate the two-loop effective potential of the non-local NJL model derived from the Sakai-Sugimoto model in string theory. In contrast to conventional NJL with 4-fermion contact interaction, the chiral symmetry was previously found to be dynamically broken for arbitrary weak coupling at the one-loop level.  As a confirmation, the approximate numerical solutions to the gap equation at one loop are explicitly demonstrated for weak couplings.  We then calculate the one and two-loop contribution to the effective potential of the non-local NJL model and found that the two-loop contribution is negative.  The two-loop potential for the chiral symmetric vacuum is also negative but larger than the combined effective potential of the chiral broken vacuum at the two-loop level.  The chiral symmetry breaking thus persists for arbitrary weak coupling at the two-loop level. 
\end{abstract}


%\arxivnumber{13xx.xxxx}

%%%%%%%%%%%%%%%%%%%%%%%%%%%%%%%%%%%%%%%
\maketitle

\section{Introduction}

Spontaneous Symmetry breaking (SSB) plays an important role in modern particle physics theory.  Higgs mechanism in the standard model, for example, can be used to generate masses of elementary particles, leptons and quarks.  The generation of quark masses by spontaneous symmetry breaking inevitably breaks the chiral symmetry of the QCD.  Chiral symmetry breaking~($\chi$SB) could also be generated {\it dynamically} by the vacuum expectation value~(vev) of chiral condensate $\psi(x) \overline{\psi}(y)$.  Dynamical $\chi$SB can explain masses of mesons and hadrons which are responsible for most of the visible mass in the universe.  It can also explain hadronic particle generation in strong interaction at low energies.  The key idea of SSB is the following.  In a theory where the Lagrangian is invariant under some symmetry while the vacuum state is not, the vacuum of the theory can generically carry non-trivial quantum number associated with the symmetry.  The theory with such vacuum is said to be in a spontaneous broken phase.  The theory could also be in a symmetric phase if its vacuum is invariant under the associated symmetry.  In the spontaneous broken phase, there is an existence of Nambu-Goldstone (NG)-boson \cite{Goldstone:1961eq}.  One can classify NG-boson into 2 cases i.e. on one hand, NG-boson is massless elementary particle and on the other hand, such boson could be a composite particle.  Dynamical symmetry breaking (DSB) usually occurs as a result of the interaction between constituent particles in the theory and yields a composite NG boson.

For chiral symmetry breaking of the QCD, the NG boson is usually identified with e.g. the three pions from the breaking of $SU(2)_{L}\times SU(2)_{R}$ to $SU(2)_{V}$ or the eight light mesons from the breaking of $SU(3)_{L}\times SU(3)_{R}$ to $SU(3)_{V}$ flavour diagonal.  This symmetry breaking pattern \cite{Vafa:1983tf, Bernard:1987sg} was successfully used to explore properties of the light hadrons and gives precise predictions of light hadronic spectra \cite{Beringer:1900zz}.  Early stage of $\chi$SB in the strong interaction was demonstrated by the linear-sigma model \cite{GellMann:1960np} and the current-algebra approach~\cite{Goldberger:1958tr,GellMann:1968rz}.  At the present, there is an incorporation between $\chi$SB and principle of effective field theory which gives a systematic framework to study QCD at low-energies, the so-called chiral perturbation theory~\cite{Gasser:1983yg}.  The theory starts with an effective theory of hadrons with chiral symmetry in the action and use the SSB to generate a chiral symmetry breaking vacuum.  The observed meson spectra shows good agreement with the prediction of the chiral perturbation theory~\cite{Scherer:2002tk,Leutwyler:1993iq}.

To address the chiral symmetry breaking/restoration phase transition, ones need to work with the action of quarks instead of hadrons.  Nambu-Jona-Lasinio~(NJL) model \cite{Nambu:1961tp} is a model of quarks with four-fermion interaction employed to demonstrate the dynamical chiral symmetry breaking in the strong interaction independent of the confinement.  Originally, NJL was formulated to explain mass of the nucleon as a consequence of the $\chi$SB.  Variations of the NJL model have been widely used as effective description of low-energy models of hadrons in QCD at zero and finite-temperature \cite{Vogl:1991qt,Klevansky:1992qe,Hatsuda:1994pi,Buballa:2003qv}.  It is also applied to break electroweak symmetry via top-quark condensation or other fermions within and beyond the standard model \cite{Miransky:1993,Bardeen:1989ds,Hill:2002ap}.  Generically, the NJL model is a very successful effective model to describe many hadronic properties in low-energy QCD, e.g. the mesons and baryons mass spectra, the pion decay constant, and the pion form factor~(see \cite{Vogl:1991qt,Klevansky:1992qe,Hatsuda:1994pi} for review).

Despite the success of the NJL as a low-energy phenomenological model approach to low-energy QCD, the original NJL model does not address confinement~(e.g. the non-confining $q\bar{q}$ discontinuities in the 2-point Green function have to be removed by introducing additional local operators~\cite{Peris:1998nj}).
There are extensions of the NJL where inclusions of non-local interactions have been proposed in the literature (see \cite{Ripka:1997} and references therein).  One can simply reproduce the non-local NJL interaction from the QCD Lagrangian by integrating out the gluon field from the one-gluon exchanging diagram \cite{Miransky:1993,Ripka:1997}.

In the non-local NJL approach, interaction depends on the momenta carried by the quarks leading to a momentum-dependent quark mass, generated by the spontaneous $\chi$SB.  It has been shown that a non-local NJL model could lead to quark confinement with acceptable values of the parameters \cite{Bowler:1994ir}.  This phenomenon originates from the fact the quark propagator has no real poles and consequently quarks have no asymptotic states.  There are several other advantages of the non-local NJL approach over the original (local) NJL model i.e. the nonlocality regularizes the model in a manner that anomalies \cite{RuizArriola:1998zi} and gauge invariance \cite{Golli:1998rf} are preserved and the momentum-dependent regulator makes the theory finite to all orders in the $1/N_c$ expansion.  Finally the dynamical quark mass is momentum dependent in contrast to the original NJL model and consistent with lattice simulations of QCD \cite{Parappilly:2005ei}.  One can see that the non-local NJL model may have more predictive power and be more realistic.  There are two major applications of the non-local NJL model in the strong interaction.  Firstly, it is incorporated in the quark model to give mass spectra of excited mesons in good agreement with the experimental data~\cite{Plant:1997jr}.  Secondly, the thermodynamics of nuclear matter and QCD phase diagram could be explained quantitatively well by using non-local NJL model~(with Polyakov-loop) \cite{General:2000zx}.

A non-local NJL model can also be constructed from certain intersecting-branes configurations in string theory.  The Sakai-Sugimoto model~(SS)~\cite{Sakai:2004cn, Sakai:2005yt} is a D8-$\overline{\rm D8}$-D4 intersecting-branes model in type IIA string theory.  The background spacetime is generated from a stack of $N_{c}$ D4-branes.  An $x^{4}$ coordinate is compactified into a circle with radius $R$ and the D4-branes wrap around the $x^{4}$.  On the boundary of the 10-dimensional space, a stack of $N_{f}$ D8 and $\overline{\rm D8}$ are located at $x^{4}=-L/2$ and $L/2$ respectively.  The left~(right)-handed quarks live on the D8~($\overline{\rm D8}$)-D4 intersection in the form of open-string excitations.  They are thus separated by distance $L$ on the boundary and there is a $U(N_{f})_{L}\times U(N_{f})_{R}$ chiral symmetry.  Geometrically, when the D8 and $\overline{\rm D8}$ merge at certain radial coordinate, the chiral symmetry breaking $U(N_{f})_{L}\times U(N_{f})_{R} \to U(N_{f})_{V}$ occurs.

We will not be considering the SS model in the full details here but would rather focus on the low-energy effective 5-dimensional field theory limit of the model~(following the work by Antonyan et. al. in Ref.~\cite{Antonyan:2006vw}), which we call the AHJK model.  In contrast to the strong coupling regime where the supergravity picture of intersecting branes provide simple geometrical interpretation of the theory, the weak coupling limit has its own unique picture of chiral symmetry breaking in terms of non-local NJL model in 5 dimensions.

 In such intersecting branes setting there are two crucial parameters i.e. the 5-dimensional 't Hooft coupling, $\lambda$ and the length scale of separation between D8-$\overline{\rm D8}$ flavor branes, $L$.  One can consider the hierarchy of those parameter as $\lambda\ll L$ which is the weak coupling regime.  In such limit, we can treat left- and right-handed quarks as weakly interacting by single (five dimensional) gluon exchange process.  The non-local NJL interaction is reproduced by integrating out gluon fields in the bulk spacetime from such D-branes configuration.  In terms of effective potential in holographic non-local NJL, the nonzero solution of chiral quark condensate exists at arbitrary weak coupling \cite{Antonyan:2006vw,Harvey:2007zz}.  In contrast, if one considers the SS model in the compactified case i.e. $R$ is finite, and includes the KK tower of states.  The $\chi$SB will happen only above a certain value of 't Hooft coupling \cite{Dhar:2009gf}.  In any cases, the analysis has been done on the effective potential of the non-local NJL at the one-loop level.  It is interesting to investigate whether the two-loop contribution would change the profile of the effective potential in any significant way.

We will start by reviewing the method of effective action in 5 dimensions when gauge fields propagate in 5 dimensions and fermions are localized in 4 dimensional subspace.  By integrating out gauge fields, we will obtain the effective fermionic action of the SS NJL model.  Subsequently, by using auxiliary field approach, we integrate out the residual fermionic fields to obtain the effective scalar action of the SS NJL model.  The gap equation is derived at one and two loop level.  At one loop, we numerically solve the gap equation for weak couplings and demonstrate the existence of the chiral broken solutions.  One-loop and two-loop contributions of the action are then calculated and discussed.  Chiral symmetry breaking is demonstrated at both one and two-loop levels.

\section{The effective Lagrangian} \label{effL}

We start with the effective action of the single-intersection model where left-handed quarks are located at a single intersection of $N_{c}$ D4 and $N_{f}$ D8 branes~\cite{Antonyan:2006vw},
\begin{eqnarray}
\mathscr{S} &=&\int d^5\,x\,\left\{-\,\frac{1}{4\,g_5^2}\,F_{MN}\,F^{MN} + \delta\,(x^4)\,q_L^\dagger\,\bar \sigma^\mu\,(i\,\partial_\mu + A_\mu)\,q_L\right\}, \label{action}
\end{eqnarray}
where $M,N,\cdots=0,1,2,3,4$ and $\mu\,\,\nu\,,\,\cdots=0,1,2,3$\,.  Integrating by part and fix the gauge, the action can be rewritten in the following form,
\begin{eqnarray}
\mathscr{S}&=& \frac{1}{g_5^2}\,\int d^5\,x\,\left\{\frac{1}{2}\,A_M\,\Box\,A^M + \delta\,(x^4)\,J^M\,A_M\right\} +  \int d^4 x\,q_L^\dagger\,\bar \sigma^\mu\,i\,\partial_\mu\,q_L, \label{act1}
\end{eqnarray}
where we have defined $J^\mu=g_5^2\,q_L^\dagger\,\bar \sigma^\mu\,q_L$\ and set $J^{(4)}=0$.  The stringy corrections are neglected in eqn.~(\ref{action}) since we take the field theory limit of the model in the weak coupling regime.  Moreover, in going from eqn.~(\ref{action}) to (\ref{act1}), we also neglected the nonlinear interactions of the gauge fields.  This approximation is justified when the distances involved are large comparing to the string length scale $\ell_{s}$ and the coupling $g_{s}N_{c}$ is small.  For sufficiently small coupling, the 't Hooft coupling $\lambda = g_{s}\ell_{s}N_{c}=g_{5}^{2}N_{c}/4\pi^{2}$ is smaller than $\ell_{s}$ and thus $\lambda\ll l_{s}\ll L$.  At the distances much larger than $\ell_{s}$, we can therefore ignore the nonlinear interactions of the gauge fields which only become significant around the distance scale $\lambda$.

The gauge fields live in 5 dimensions and it is natural to integrate them out to obtain 4-dimensional effective action of the fermions.  For consideration of the chiral symmetry breaking, we can bosonize the fermion bilinear and integrate out the fermions subsequently.  In order to integrate out the gauge field $A_M$.
We recall the procedure from \cite{Donoghue:1992dd}, start with
\begin{eqnarray}
e^{i\int\,d^4x\,\mathscr{L}_{\rm eff}} = \int\,[d\,H]\, e^{i\,\int\,d^4x\,\mathscr{L}\big(H(x),l(x)\big)}\,\Big / \int\,[d\,H]\, e^{i\,\int\,d^4x\,\mathscr{L}\big(H(x),0\big)}, \label{int-out}
\end{eqnarray}
where $H(x)$\, and $l(x)$ are heavy and light fields respectively. In our case, $A_M$ is the heavy field and $q_{L,R}$ are the light fields. The actions with and without the light fields are given by
\begin{eqnarray}
\int d^5\,x\,\mathscr{L}\big(A_M,q_L\big) &=& \frac{1}{g_5^2}\,\int d^5\,x\,\left\{\frac{1}{2}\,A_M\,\Box\,A^M
+ \delta\,(x^4)\,J^M\,A_M\right\}
\nonumber\\
&\;&\quad +\,\int d^4 x\,q_L^\dagger\,\bar \sigma^\mu\,i\,\partial_\mu\,q_L\,,\label{with-H}\\
\int d^5\,x\,\mathscr{L}\big(A_M,0\big) &=& \frac{1}{g_5^2}\,\int d^5\,x\,\frac{1}{2}\,A_M\,\Box\,A^M\label{without-H}\,.
\end{eqnarray}
By using functional path integral as demonstrated in appendix~\ref{appa}, the effective action after integrating out the gauge field can be read off from eq.~(\ref{res-1}),
\begin{eqnarray}
\mathscr{S}_{\rm eff} &=& i\int d^4 x\,q_L^\dagger\,\bar \sigma^\mu\,\partial_\mu\,q_L
\nonumber\\
&\;&\qquad -\,\frac{g_5^2}{16\,\pi^2}\int\,d^4x\,\,d^4y\,G(x-y\,,\,0)\,\Big[q_L^\dagger(x)\,\bar \sigma^\mu\,q_L(y)\Big]\,\Big[q_L^\dagger(y)\,\bar \sigma_\mu\,q_L(x)\Big].
\end{eqnarray}
Next, we extend the Lagrangian (\ref{action}) to the the left and right-handed quark fields located at different intersections, D4-D8 and D4-$\overline{\rm D8}$ respectively~\cite{Antonyan:2006vw}\,, this is the low-energy field theory limit of the SS model,
\allowdisplaybreaks[2]
\begin{eqnarray}\label{actionfull}
\mathscr{S} &=& \int d^5\,x\,\Bigg\{-\,\frac{1}{4\,g_5^2}\,F_{MN}\,F^{MN} + \delta\left(x^4 + \frac{L}{2} \right)\,q_L^\dagger\,\bar \sigma^\mu\,(i\,\partial_\mu + A_\mu)\,q_L
\nonumber\\
&\;&\qquad\qquad\qquad\qquad\qquad\qquad\; +\,\delta\left(x^4 - \frac{L}{2} \right)\,q_R^\dagger\,\sigma^\mu\,(i\,\partial_\mu + A_\mu)\,q_R\Bigg\},\nonumber\\
&=& \frac{1}{g_5^2}\int d^5\,x\,\Bigg\{A_M^{(L)}\,g^{MN}\,\Box\,A_N^{(R)} + \delta\left(x^4 + \frac{L}{2} \right)\,J_{(L)}^M\,A_M^{(L)} + \delta\left(x^4 - \frac{L}{2} \right)\,J_{(R)}^M\,A_M^{(R)}\Bigg\}\nonumber\\
&\;&\qquad\qquad\qquad\qquad + \int d^4 x\,q_L^\dagger\,\bar \sigma^\mu\,i\,\partial_\mu\,q_L + \int d^4 x\,q_R^\dagger\,\sigma^\mu\,i\,\partial_\mu\,q_R  \label{act2}
\end{eqnarray}
where we define \,$J^M_{(L)}\equiv g_5^2\,q_{L}^\dagger\,\bar\sigma^M\,q_{L}$\,, \,$J^M_{(R)}\equiv g_5^2\,q_{R}^\dagger\,\sigma^M\,q_{R}$\, and \,$A_M^{(L)}$\,, \,$A_M^{(R)}$\, are the gauge fields in 5-dimensional spacetime which are located on the D4-D8 and D4-$\overline{\rm D8}$ intersections respectively.  As before when we obtained (\ref{act1}), the nonlinear interactions of the gauge fields are negligible if the distances involved are larger than the string scale $\ell_{s}$ for $\lambda \ll \ell_{s} \ll L$, i.e. when the coupling is weak.  We therefore ignored these interactions in (\ref{act2}).  \\

Then the generating functional of the above action is given by
\begin{eqnarray}
&\;&\int [d\,A_M^{(L)}\,d\,A_M^{(R)}]\,\Delta_{FP}\,\exp\left\{i\,\mathscr{S}\big(A_M^{(L)},A_M^{(R)},q_L,q_R\big)\right\}\nonumber\\
&\;& = \int [d\,A_M^{(L)}\,d\,A_M^{(R)}]\,\Delta_{FP}
\nonumber\\
&\;&\times\,\exp\,\Bigg\{\,\frac{i}{g_5^2}\,\int\,d^5x\,A_M^{(L)}\,g^{MN}\,\Box\,A_N^{(R)}\nonumber\\
&\;&\qquad\qquad
-\,\frac{i}{g_5^2}\int\,d^5x\,\,d^5y\,\delta\left(x^4 + \frac{L}{2}\right)\delta\left(y^4 - \frac{L}{2}\right)J_{(L)}^M(x)\,G_{MN}(x-y\,,\,x^4-y^4)\,J_{(R)}^N(y) \nonumber\\
&\;&\qquad\qquad\qquad\qquad\qquad\qquad +\, i \int d^4 x\,q_L^\dagger\,\bar \sigma^\mu\,i\,\partial_\mu\,q_L + i \int d^4 x\,q_R^\dagger\,\sigma^\mu\,i\,\partial_\mu\,q_R \Bigg\}. \label{path-HLR}
\end{eqnarray}
Using eq.~(\ref{path-HLR}), the effective action in the integrating out procedure is written by
\begin{eqnarray}
e^{i\,\mathscr{S}_{\rm eff}}  &=&\frac{ \int [d\,A_M^{(L)}\,d\,A_M^{(R)}]\,\Delta_{FP}\,\exp\left\{i\,\mathscr{S}\big(A_M^{(L)},A_M^{(R)},q_L,q_R\big)\right\}}{\int [d\,A_M^{(L)}\,d\,A_M^{(R)}]\,\Delta_{FP}\,\exp\left\{i\,\mathscr{S}\big(A_M^{(L)},A_M^{(R)},0,0)\right\}}\,,
\nonumber\\
&=& \exp\,\Bigg\{i \int d^4 x\,(q_L^\dagger\,\bar \sigma^\mu\,i\,\partial_\mu\,q_L + q_R^\dagger\,\sigma^\mu\,i\,\partial_\mu\,q_R)
\nonumber\\
&\;&\qquad\quad\, -\,i\,g_5^2\int\,d^4x\,\,d^4y\,q_{L}^\dagger(x)\,\bar\sigma^\mu\,q_{L}(x)\,G_{\mu\nu}\left(x-y\,,\,L\right)\,q_{R}^\dagger(y)\,\sigma^\nu\,q_{R}(y) \,\Bigg\}.
\end{eqnarray}

Finally, we obtain the effective non-local Lagrangian in the Feynman gauge as
\begin{eqnarray}
\mathscr{S}_{\rm eff} &=& \int d^4 x\,(q_L^\dagger\,\bar \sigma^\mu\,i\,\partial_\mu\,q_L + q_R^\dagger\,\sigma^\mu\,i\,\partial_\mu\,q_R)
\nonumber\\
&\;&-\,g_5^2\int\,d^4x\,\,d^4y\,\frac{1}{8\,\pi^2}\,g_{\mu\nu}\,G\left(x-y\,,\,L\right)q_{L}^\dagger(x)\,\bar\sigma^\mu\,q_{L}(x)\,\,q_{R}^\dagger(y)\,\sigma^\nu\,q_{R}(y),
\nonumber\\
&=& \int d^4 x\,(q_L^\dagger\,\bar \sigma^\mu\,i\,\partial_\mu\,q_L + q_R^\dagger\,\sigma^\mu\,i\,\partial_\mu\,q_R)
\nonumber\\
&\;& +\,\frac{g_5^2}{4\,\pi^2}\int\,d^4x\,\,d^4y\,G\left(x-y\,,\,L\right)[\,q_{L}^\dagger(x)\cdot\,q_{R}(y)]\,[\,q_{R}^\dagger(y)\cdot\,q_{L}(x)],
\label{eff-action}
\end{eqnarray}
where we used the Fierz identity \,$\left( q_L^\dagger(x)\,\bar\sigma^\mu\,q_L(x)\right)\left( q_R^\dagger(y)\,\sigma_\mu\,q_R(y)\right)=-\,2\left( q_L^\dagger(x)\cdot q_R(y)\right)$ $\left( q_R^\dagger(y)\cdot q_L(x)\right)$.  The dot in the right-hand side is the contraction in the colour indices, therefore each fermion bilinear in the final expression of the effective interaction Lagrangian is a colour singlet.  There is a non-local interaction between two colour singlet operators in the theory.

\section{Effective potential at one-loop : Auxiliary field approach}
In this section, we will calculate the effective potential from the effective action eq.~(\ref{eff-action}).  We will use the standard method of effective field theory i.e. bosonize the fermion bilinear which would become the chiral condensate and integrate out the heavy-residual fields (in our case is the fermion fields).  The effective potential with one-loop radiative correction can be obtained subsequently from the effective action.

Following ref.~\cite{Antonyan:2006vw}, we start with the auxiliary field method.  This method is used to study the symmetry breaking of the model by introducing the auxiliary field to the effective Lagrangian. In our case is the bosonized complex fields i.e.
\begin{eqnarray}
T(x,y) &=& \frac{\lambda}{N_c}\,G(x-y,L)\,q_L^\dagger(x)\cdot\,q_R(y),
\nonumber\\
\bar T(y,x) &=&  T^\dagger(x,y) =\frac{\lambda}{N_c}\,G(x-y,L)\,q_R^\dagger(y)\cdot\,q_L(x),
\label{aux-field}
\end{eqnarray}
where the coupling $\lambda/N_c$ is related to the $g_5^2$ coupling in the effective Lagrangian by the relation $\lambda = \textstyle{\frac{g_5^2}{4\,\pi^2}}\,N_c$\,.

Substituting auxiliary fields from eq.~(\ref{aux-field}) into the effective action eq.~(\ref{eff-action}), we obtain
\begin{eqnarray}
\mathscr{S}_{\rm eff} &=& \int d^4 x\,(q_L^\dagger\,\bar \sigma^\mu\,i\,\partial_\mu\,q_L + q_R^\dagger\,\sigma^\mu\,i\,\partial_\mu\,q_R)
\\
&+& \int\,d^4x\,\,d^4y\,
\left( -\,\frac{N_c}{\lambda}\,\frac{T(x,y)\,\bar T(x,y)}{G\left(x-y\,,\,L\right)}
+ \bar T(y,x)\,q_{L}^\dagger(x)\cdot\,q_{R}(y) + T(x,y)\,q_{R}^\dagger(y)\cdot\,q_{L}(x)\right).
\nonumber
\label{aux-action}
\end{eqnarray}
In the chiral~(Weyl) basis, one can rewrite the Lagrangian as
\begin{eqnarray}
\mathscr{S}_{\rm eff} &=& \int d^4x\,\bar q(x)\,\Big( i\,\partial\!\!\!/ + T(x)\,P_L + \bar T(x)\,P_R \Big)\,q(x)
- \int\,d^4x\,
\frac{N_c}{\lambda}\,\frac{T(x)\,\bar T(x)}{G\left(x,\,L\right)},
\label{aux-action-chiral}
\end{eqnarray}
where we imposed the simplifying ansatz $T(x,y)=T(|x-y|)$ consistent with the Poincare symmetry of the expectation value of the operator.  This is justified since we are considering expectation value of $T(x,y)$ in the vacuum to study the chiral symmetry breaking.
\begin{figure}[]
\noindent
\begin{center}
\includegraphics[width=10cm,clip=true]{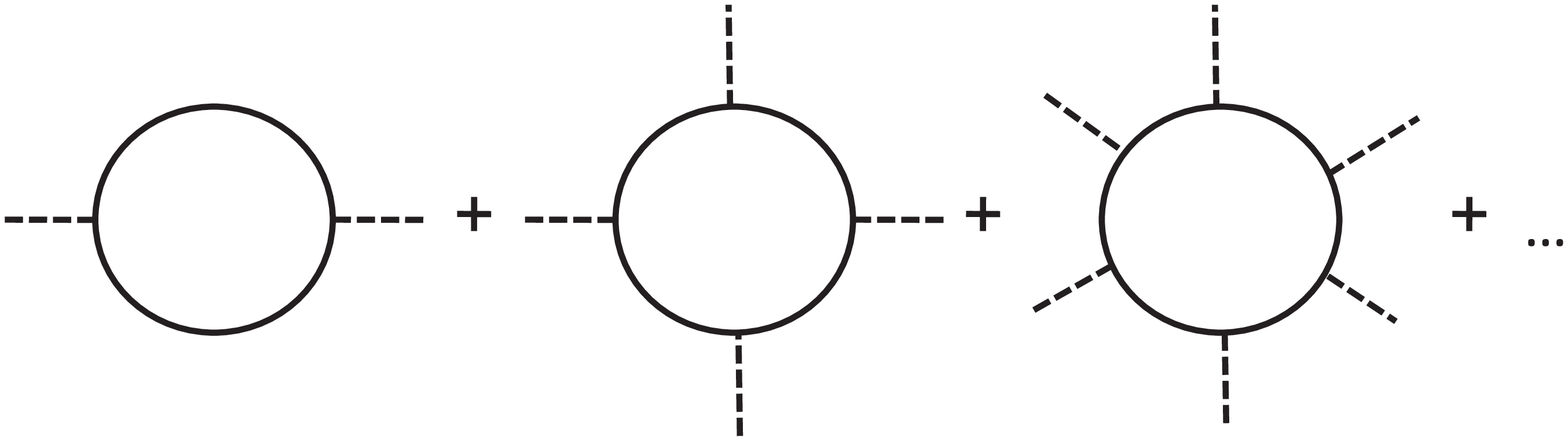}
\end{center}
\caption{One-loop expansion of fermion fields.}
\label{one-loop}
\end{figure}

We are ready to integrate out the fermion fields in eq.~(\ref{aux-action-chiral}), it reads
\begin{eqnarray}
e^{i\,\mathscr{S}_{\rm eff}}  &=& \int [d\bar q\, d q ]\,\exp\left\{i\,\mathscr{S}\big( \bar q, q, T, \bar T \big)\right\}\,\Bigg / \int [d\bar q\, d q ]\,\exp\left\{i\,\mathscr{S}\big( \bar q, q, 0, 0\big)\right\},
\nonumber\\
&=& \exp\,\Bigg\{{\rm Tr}\,\ln\left( 1 + \frac{T(x)\,P_L + \bar T(x)\,P_R}{i\,\partial\!\!\!/} \right)- \,i\,\int\,d^4x\,\frac{N_c}{\lambda}\,\frac{T(x)\,\bar T(x)}{G\left(x,\,L\right)}\,\Bigg\}.
\end{eqnarray}
The identities $\int [d\bar q\, d q ]\,\exp\left\{i\int d^4x\,\bar q(x)\,\mathcal{A}\,\,q(x)\right\} = \det \mathcal{A} = \exp({\rm Tr}\,\ln \mathcal{A})$\, are used above.  Then the effective potential with one-loop expansion can be determined from the effective action,
\begin{eqnarray}
V_{\rm eff} &=& -\,\mathscr{S}_{\rm eff}
\nonumber\\
&=&  \int\,d^4x\,\frac{N_c}{\lambda}\,\frac{T(x)\,\bar T(x)}{G\left(x,\,L\right)} +\,i\,{\rm Tr}\,\ln\left( 1 + \frac{T(x)\,P_L + \bar T(x)\,P_R}{i\,\partial\!\!\!/} \right),
\end{eqnarray}
where ${\rm Tr} \equiv {\rm Tr}_{\rm spinor}\,{\rm Tr}_{\rm colour}\,{\rm Tr}_{\rm flavor}\,{\rm Tr}_{\rm spacetime}\,$ is the trace over all indices (i.e. spinor, color, flavor, spacetime). The physical meaning of this procedure is depicted by figure \ref{one-loop}.\\

The second term in the effective potential can be calculated by expansion
\allowdisplaybreaks[1]
\begin{eqnarray}
{\rm Tr}\,\ln\left( 1 + \frac{(T\,P_L + \bar T\,P_R)}{i\,\partial\!\!\!/} \right)
&=&{\rm Tr}_{\rm spinor}\,{\rm Tr}_{\rm color}\,{\rm Tr}_{\rm flavor}\,{\rm Tr}_{\rm spacetime}
\nonumber\\
&\;&\times\,\sum_{n=1}^{\infty}\,\frac{(-1)^{n-1}}{n}
\left[\frac{(T\,P_L + \bar T\,P_R)}{i\,\partial\!\!\!/}\right]^n,
\nonumber\\
&=& i \,N_c\,N_f\,V\,\int\,\frac{d^4k_E}{(2\,\pi)^4}\,\ln
\left(1 + \frac{T(k_E)\,\bar T(k_E)}{k_E^2}\right),
\end{eqnarray}
where we used the following relations;
\begin{eqnarray}
 {\rm Tr}_{\rm spinor}\,\mathbf{1}_{\rm spinor}&=& 2\;({\rm in~chiral~basis})\,,\quad {\rm Tr}_{\rm color}\,\mathbf{1}_{\rm color} = N_c\,,
\nonumber\\
{\rm Tr}_{\rm flavor}\,\mathbf{1}_{\rm flavor} &=& N_f\,,\quad{\rm Tr}_{\rm spacetime} = \int d^4x = V\,,
\nonumber\\
\frac{1}{\partial\!\!\!/ } &=& \int \frac{d^4k}{(2\,\pi)^4}\,\frac{k\!\!\!/}{k^2}\,e^{i\,k\cdot\,(x-y)}\,,\qquad
 (T\,k\!\!\!/\,P_L + \bar T\,k\!\!\!/\,P_R)^2  =  T\,\bar T\,k^2.
\end{eqnarray}
The momentum has been Euclideanized and henceforth we will drop the subscript.

Finally, the effective potential at one-loop is given by (scaled by factor $N_f$)
\begin{eqnarray}
V_{\rm 1-loop} &=&  N_{c}\left[\int\,d^4x\,T(x)\,\bar T(x)\,\frac{(x^2 + L^2)^{\frac{3}{2}}}{\lambda} -\int\,\frac{d^4k}{(2\,\pi)^4}\,\ln
\left(1 + \frac{T(k)\,\bar T(k)}{k^2}\right)\right].  \label{V1}
\end{eqnarray}

The equation of motion~(the gap equation) of the scalar $T(x)$ from the effective action, eq.~(\ref{V1}), is
\begin{eqnarray}
 \int\,d^4x\,T(x)\,e^{-ikx}\frac{(x^2 +L^2)^{\frac{3}{2}}}{\lambda}& = & \frac{T(k)}{k^{2}+T(k)\,\bar T(k)}. \label{gap}
\end{eqnarray}
Apart from the trivial solution $T = 0$ for the chiral-symmetric vacuum, the general solution to the gap equation $\delta V_{\rm eff}/\delta \bar{T}(k)=0$ can be solved perturbatively either analytically or numerically~(see appendix \ref{gapeq}).  Non-vanishing $T$ solution corresponds to chiral symmetry breaking vacuum which has lower energy and thus represents a true vacuum.  We can obtain approximate solution by solving the gap equation in 2 regions of momentum, small and large $k$~(i.e. $T(k)\bar{T}(k)\gg k^{2}$ and $T(k)\bar{T}(k)\ll k^{2}$ respectively).  The two solutions then can be matched to determine the unknown constants.  An approximate solution from such method is in the following form~\cite{Antonyan:2006vw}
\begin{eqnarray}
T(k)=\bar T(k) =
\begin{cases}
\;T_{0}=k_*\equiv\sqrt{\frac{\lambda}{L^3}} & ,\; 0\,<\,k\,\leq\,k_*\,,
\\
\\
\;T_{0}^2\,\frac{e^{-\,L\,k}}{k}\equiv\frac{\lambda}{L^3}\,\frac{e^{-\,L\,k}}{k} & ,\; k_*\,<\,k\,<\,\Lambda\,.
\end{cases}  \label{Gfun}
\end{eqnarray}

Generically by using the gap equation, the one-loop potential can be rewritten to be
\begin{eqnarray}
V_{\rm 1-loop} & = &  N_{c}\int\,\frac{d^4k_E}{(2\,\pi)^4}\, \Bigg[ \frac{T(k)\bar{T}(k)}{k^{2}+T(k)\bar{T}(k)} - \ln
\left(1 + \frac{T(k)\,\bar T(k)}{k^2}\right)\Bigg].  \label{Vone}
\end{eqnarray}
By substituting approximate propagator eq.~(\ref{Gfun}) into eq.~(\ref{Vone}), we can demonstrate that there is chiral symmetry breaking vacuum induced by small momentum contribution to the one-loop potential.  The details are discussed in section \ref{disc}.  Essentially, since the integrand in eq.~(\ref{Vone}) is a negative-definite function of variable $k^{2}/T(k)\bar{T}(k)$, the one-loop potential is always negative for nonzero $T$ regardless of the exact form of the solution of the gap equation.  It is obvious that the solution with nonzero $T$ gives the lower potential than the chiral symmetric solution $T=0$.

It is remarkable that the chiral symmetry breaking of the one-loop potential occurs at any weak coupling.  The reasons are the boundness of the positive classical term~(the first term in the right-hand side of eq.~(\ref{Vone})) whilst the negative loop term~(the second term in the right-hand side of eq.~(\ref{Vone})) is not bounded for low momentum.  The solution of the gap equation, eq.~(\ref{Gfun}), is a constant for the low momentum, resulting in $\ln (1/k^{2})$-divergence of the loop term as $k\to 0$, regardless of $\lambda$.

%%%%%%%%%%%%%%%%%%%%%%%%%%%%%%%%%%%%%%%%%%%%%%%%%%%%%%%%%%%%%%%%%%%%%%%%%%%%%%%%%%%%%%%%%%%%%%%%%%%%%%%%%%%%%%%%%%%%%%%%%%%%%%%%%%%%%%%%%%%%%%%%%%%%%%%%%%%%%%%%
\section{Numerical solutions to the gap equation at one-loop level}

Before we proceed to the evaluation of two-loop contribution to the effective potential, we will demonstrate that the gap equation, eq.~(\ref{gap}), has actual solution for arbitrary weak coupling.  As stated in Ref.~\cite{Antonyan:2006pg, Antonyan:2006qy}, the Green function given by eq.~(\ref{propa}), $G(x,L)=(x^{2}+L^{2})^{-3/2}$, is a long range interaction in 4 dimensions as we can see from the divergence of the integral
\begin{eqnarray}   
\int G(x,L)~d^{4}x.
\end{eqnarray}
This is a generic feature of the Lorentzian $(1+d)$-propagator originated in higher dimension when projected onto lower Euclidean $d$-dimension.  The nonlocal Green function is the result of one-gluon exchange interaction in 5 dimensions which would become short range only in $2$ and lower dimensions.  Interestingly when projected onto $2$ dimensions, the model becomes a nonlocal generalization of the Gross-Neveu model~\cite{Antonyan:2006qy, Gross:1974jv} which breaks chiral symmetry at any coupling.  

We solve the gap equation by the procedure used in Ref.~\cite{Dhar:2009gf}.  First we define 
\begin{eqnarray}
\phi(x) & \equiv &  \frac{1}{N_{c}}\langle q_L^\dagger(x)\cdot\,q_R(0)\rangle = \frac{\phi_{0}}{4\pi^{2}l^{3}}\varphi(x/l),
\end{eqnarray}
with $l$ being the chiral symmetry breaking length scale and $\phi_{0}$ is a constant.  Substitute into the gap equation, we obtain
\begin{eqnarray}     
f(p) & = & \frac{\bar{\lambda}t(p)}{p^{2}+\bar{\lambda}^{2}\phi_{0}^{2}t^{2}(p)}, \label{pgap}
\end{eqnarray}
where $\bar{\lambda}\equiv \lambda l^{2}$ and 
\begin{eqnarray}
f(p)&\equiv&\frac{1}{p}\int_{0}^{\infty}~J_{1}(py)\varphi(y)~y^{2}~dy, \\
t(p)&\equiv&\frac{1}{p}\int_{0}^{\infty}~J_{1}(py)\varphi(y) G(ly, L)~y^{2}~dy.
\end{eqnarray}
The Fourier transform $\phi(k), T(k)$ are related to $f(p), t(p)$ by $\phi(k)=\phi_{0}lf(p), T(k)=\lambda\phi_{0}l t(p)$ where $p=kl$.  The numerical solution to the gap equation can be obtained by finding the trial function for $\varphi(x)$ which satisfies eq.~(\ref{pgap}).  This could be done by adjusting the parameters of the trial function such that they minimize the quantity
\begin{eqnarray}  
(\delta f)^{2}\equiv \int_{0}^{\infty}\left( f(p)-f_{s}(p)\right)^{2}~dp, 
\end{eqnarray}
where 
\begin{eqnarray}
f_{s}(p) & \equiv & \frac{\bar{\lambda}t(p)}{p^{2}+\bar{\lambda}^{2}\phi_{0}^{2}t^{2}(p)}.
\end{eqnarray}

\subsection{AHJK solution}

We review the approximate solution derived in Ref.~\cite{Antonyan:2006vw} given by eq.~(\ref{Gfun}) which we will call the AHJK solution.   It can be shown that the Fourier transformed gap equation, eq.~(\ref{pgap}) is satisfied up to the order of $\mathcal{O}(\lambda)$ by this ansatz.  For high $k>k_{*}$, the solution is approximated by a constant condensate $\phi(x)=\phi_{0}/4\pi^{2}l^{3}=1/4\pi^{2}L^{3}$~(from eq.~(3.16) of Ref.~\cite{Antonyan:2006vw}) leading to 
\begin{eqnarray}  
\phi(k) & = & \int d^{4}x ~\phi(x) e^{-i k x} = \frac{4\pi^{2}}{L^{3}}\delta^{(4)}(k), \\
f(p) & = & \frac{4\pi^{2}}{l^{4}}\delta(k), \label{fahjk}
\end{eqnarray}
where $\phi_{0}=l^{3}/L^{3}$.  The direct integration of $t(p)$ gives
\begin{eqnarray}
t(p) & = & \frac{e^{-p L/l}}{p l^{3}}=\frac{1}{\lambda \phi_{0}l}T(k).
\end{eqnarray}
Consequently, for high $p$
\begin{eqnarray}
f_{s}(p) \simeq \frac{\bar{\lambda}t(p)}{p^{2}} =\frac{\lambda e^{-kL}}{k^{3}l^{4}},
\end{eqnarray}
which is of order $\mathcal{O}(\lambda)$ and vanishing with $k$.  Apparently, $f(p)$ from eq.~(\ref{fahjk}) becomes zero for high $k$, therefore the gap equation is satisfied up to an order of $\mathcal{O}(\lambda)$.   

For low $k<k_{*}$, $T(x)=\bar{T}(x)=T_{0} \delta^{(4)}(x)$ and $T(k)=\bar{T}(k)=T_{0}$ are the solutions to the gap equation.  Straighforward substitution gives
\begin{eqnarray}
f(p) & = & \frac{1}{T_{0}\phi_{0}l}, \quad t(p) = \frac{T_{0}}{\lambda \phi_{0}l}.
\end{eqnarray}
Consequently for $T(k)\bar{T}(k)=T_{0}^{2}\gg k^{2}$,
\begin{eqnarray}
f_{s}(p) & \simeq & \frac{\bar{\lambda}t(p)}{\phi_{0}^{2}\bar{\lambda}^{2}t^{2}(p)},
\end{eqnarray}
equal to $f(p)$ exactly in this limit regardless of $\phi_{0}$.   

In this section we have demonstrated that the Fourier transformed gap equation is satisfied by the AHJK solution given by eq.~(\ref{Gfun}) at the order of $\mathcal{O}(\lambda)$.  As long as $\lambda/L$ is small, the use of this ansatz in the evaluation of the effective potential is justified.  For completeness, we also present other classes of solutions in section \ref{L1} and \ref{L2}.  These numerical solutions are found in the region of the parameter space with $\lambda/L > 1$, where the gauge dynamics become important~(see also Ref.~\cite{Dhar:2009gf} for the same kind of solutions when the Kaluza-Klein states are included).

\subsection{solutions with $l \lesssim L$}  \label{L1}

\begin{figure}
\begin{center}
        \subfigure[]{\includegraphics[width=0.5\textwidth]{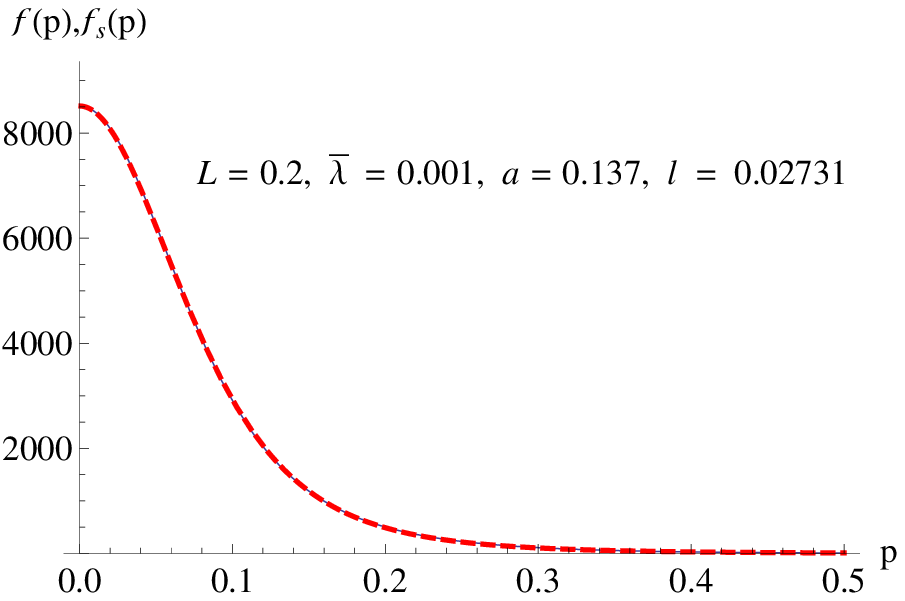}}\\
        \subfigure[]{\includegraphics[width=0.5\textwidth]{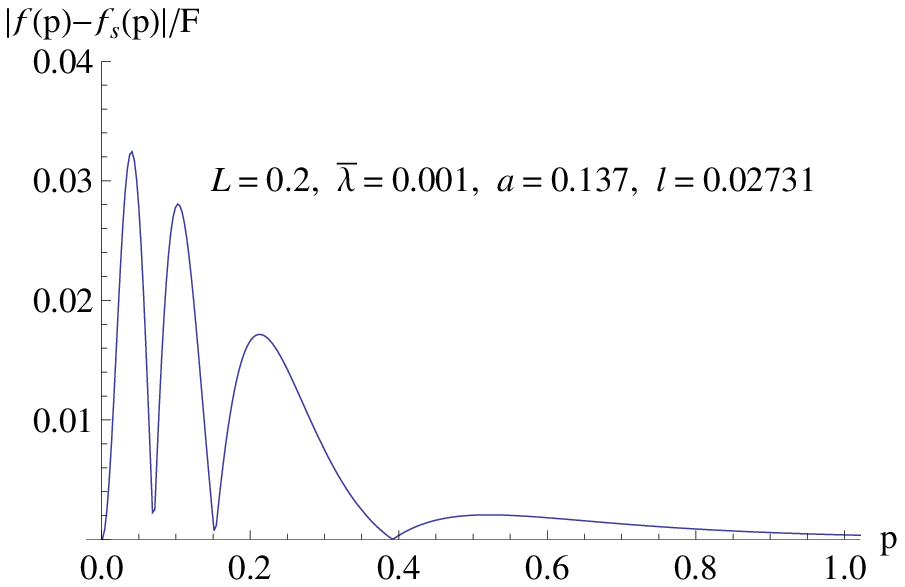}}\hfill
        \subfigure[]{\includegraphics[width=0.5\textwidth]{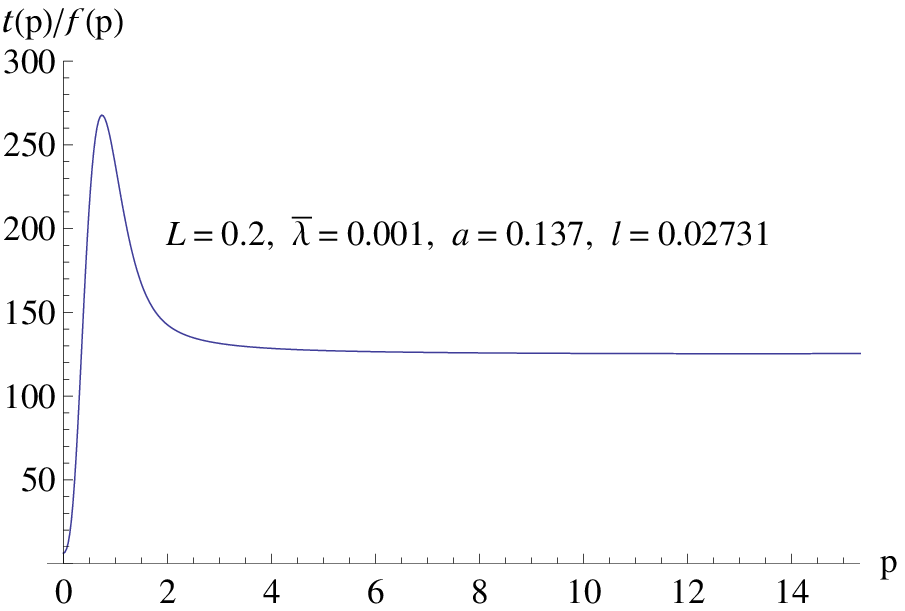}}
\end{center}
\caption{Numerical solutions to the gap equation for $\bar{\lambda} = 0.001$, $f_{s}(p)~(f(p))$ is represented by solid blue~(dashed red) line in (a).  The matching of the two functions implies the solution to the gap equation at one loop.  The error estimate is represented in (b)}\label{figgap}
\end{figure}

First, we search for solutions with the condensate scale $l$ smaller than $L$.  The trial function we use is in the exponential form
\begin{eqnarray}
\varphi(x)& = & e^{-a x}, \label{trfun}
\end{eqnarray}
with only one parameter $a$ to determine.  By adjusting two parameters, $a, l$, for a fixed $L$ and $\bar{\lambda}$, we found numerical solutions for $\bar{\lambda} = 0.001-0.1$.  The numerical solution for $\bar{\lambda}=0.001, L=0.2$ is shown in Figure~\ref{figgap}, where $F\equiv \int_{0}^{\infty}f(p) dp$.  The error estimate for this solution is $\delta f/F=0.0095$~(about 1 \%) with the momentum distribution shown in Fig.~\ref{figgap} (b).  Oscillating behaviour of the error in the momentum space is due to the Bessel function $J_{1}(py)$ in the Fourier transform.  Similar solutions exist for other values of $\bar{\lambda}$, the list of certain values are given in Table~\ref{tab1}.  

\begin{table}[h]
\caption{Approximate solutions to one-loop gap equation for $L=0.2$}\label{tab1}
\begin{center}
\begin{tabular}{|l|l|l|l|l|}
\hline 
$\quad\bar{\lambda}$&~\quad$a$&~\quad$l$&\quad $\lambda$&$~\delta f/F$   \\  \hline
0.001&0.137&0.02731&1.3408&0.0095   \\
0.01&0.4245&0.08641&1.3393&0.0056   \\
0.03&0.81&0.149&1.3513&0.0061   \\
0.05&1.0&0.193&1.3423&0.0039   \\
0.1&1.35&0.273&1.3418&0.0031    \\ \hline
\end{tabular}
\end{center}
\end{table}

It is remarkable that the various small-$l$ solutions to the one-loop gap equation appear to have the same value of $\lambda \simeq 1.34$, in a strong coupling regime.  It is interesting to investigate what happens at this coupling in the future work.      

Few comments on high-momentum bahaviour of $f(p), t(p)$ and $f_{s}(p)$ are in order.  For the trial exponential function given by eq.~(\ref{trfun}), we can integrate to obtain
\begin{eqnarray}
f(p)& = & \frac{3}{a^{4}}\left( 1+\frac{p^{2}}{a^{2}}\right)^{-5/2},
\end{eqnarray}
which approaches $p^{-5}$ dependence for large $p$.
We also found both analytically and numerically that the ratio $t(p)/f(p)$ approaches a constant $1/L^{3}$ as $p$ increases~(which is the consequence of $l\ll L$), as shown in Fig.~\ref{figgap} (c).  Consequently, $f_{s}(p)\simeq {\rm(Const.)}\bar{\lambda}a^{5}p^{-7}$ and $f(p) - f_{s}(p)\sim p^{-5}$ for large $p$.  Namely, the error of the matching vanishes very rapidly with increasing momentum.  

The appropriate interpretation is the following.  Even though the matching between $f(p)$ and $f_{s}(p)$ is excellent for low momentum, they have different $p$-dependence for high momentum.  The exponential trial function can be served as a good approximate solution to the one-loop gap equation in the low momentum region where the distance scale involved is large.  As the momentum increases beyond certain value~(around $p\simeq 1$ or momentum $k\simeq 1/l$), the one-loop gap equation no longer has condensate solution, at least in the exponential form.  However, since the UV cutoff $\Lambda \simeq 1/L < 1/l$, the high momentum region is not relevant.  As long as the large scale physics of chiral symmetry breaking is concerned~(not smaller than $l$), the trial exponential function is an excellent approximate solution to the gap equation.  

\subsection{solutions with $l > L$}  \label{L2}

We also found the class of solutions with $l > L$ of the one-loop gap equation by using the trial function
\begin{eqnarray}
\varphi(x) & = & \frac{e^{-a x}}{(1+x^{2}b^{2})^{\epsilon}},  \label{tfunc}
\end{eqnarray}
with parameters $a, b, \epsilon$ to be determined.  As suggested by Ref.~\cite{Dhar:2009gf}, given a value of $\lambda$, we choose $b=l/L$, set $l=1$ and adjust $L~(<l), a, \epsilon$ to find a matching between $f(p)$ and $f_{s}(p)$.  The chiral broken solution should exist for any value of $L$~(with corresponding value of $l$) but for convenience, we choose to fix $l=1$ and let $L$ be small quantities to be determined.  

\begin{figure}
\begin{center}
        \subfigure[]{\includegraphics[width=0.5\textwidth]{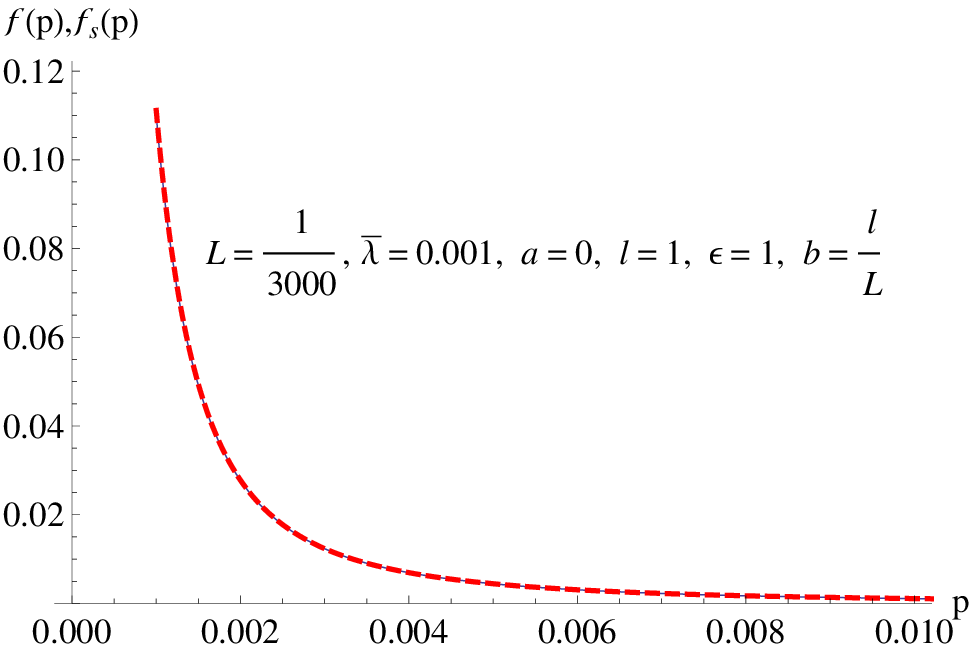}}\hfill
        \subfigure[]{\includegraphics[width=0.5\textwidth]{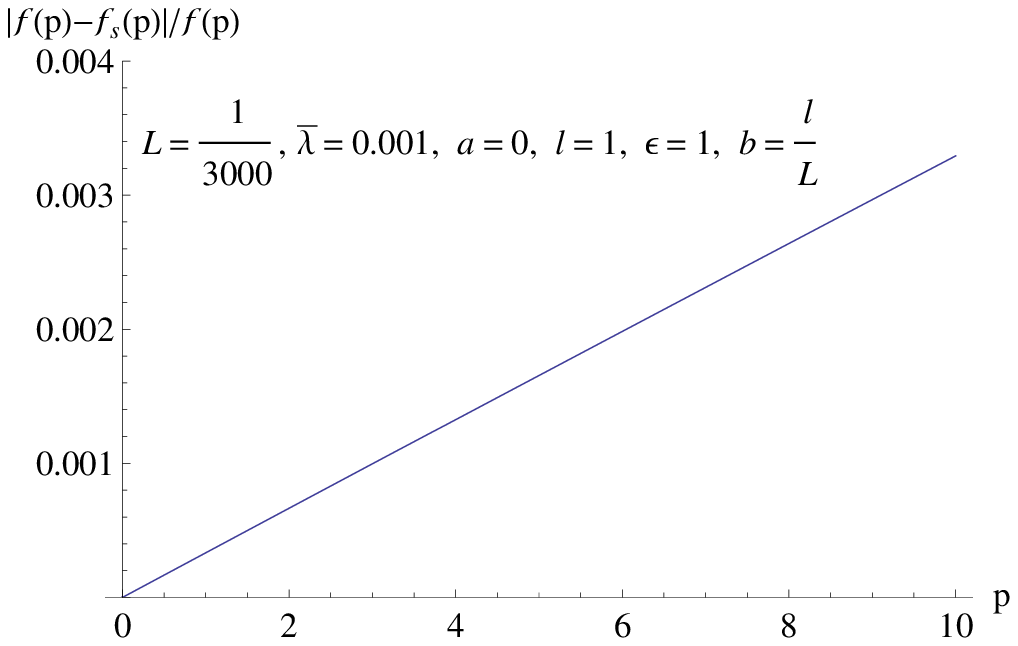}}
\end{center}
\caption{Numerical solutions to the gap equation for $l \gg L, \bar{\lambda} = 0.001$, $f_{s}(p)~(f(p))$ is represented by solid blue~(dashed red) line in (a).  The matching of the two functions implies the solution to the gap equation at one loop.  The error estimate is represented in (b)}\label{figgap1}
\end{figure}

A class of approximate solutions is found with $a=0, \epsilon = 1$ for arbitrary weak coupling.  Figure~\ref{figgap1} shows one such solution for very weak coupling $\lambda = 0.001$.  Other solutions for $\lambda =0.1, 0.01$ also exist with $L=1/30, 1/300$~(i.e. solutions with $\lambda/L =3$) respectively.  In this $a=0$ case, we can directly integrate
\begin{eqnarray}
f(p) & = & \frac{1}{p b^{3}}~K_{1}\left(\frac{p}{b}\right) \sim \sqrt{\frac{\pi}{2p^{3} b^{5}}}~e^{-p/b}, \\
t(p) & \simeq & \frac{\pi}{2 p l^{3}}\left( I_{1}\left(\frac{p}{b}\right) - \bold{L}_{1}\left(\frac{p}{b}\right)\right)~(l \gg L) \sim \frac{1}{p l^{3}},
\end{eqnarray}
for large $p$.  The function $I_{1}(x), K_{1}(x)$ are the modified Bessel function of the first and second kind and $\bold{L}_{1}(x)$ is the modified Struve function.  Apparently, the ratio
\begin{eqnarray}
\frac{f_{s}(p)}{f(p)} \sim \frac{\bar{\lambda}}{l^{3}}\sqrt{\frac{2 b^{5}}{\pi}}~p^{-3/2}e^{p/b},
\end{eqnarray}
is not equal to $1$ for large $p$.  The error $f(p)-f_{s}(p)$ diminishes with increasing $p$ as a result of vanishing $f(p), t(p)$.  Similar to the case with $l\lesssim L$, this class of approximate solutions is valid only for low momentum below the UV cutoff, $k < 1/l$.  Notably, the $a=0$ solutions have infrared divergence as $p\to 0$ and the IR cutoff is also required.  In contrast to high momentum, the matching becomes exceedingly accurate as $p$ decreases towards zero.  

Both classes of solutions, $l \lesssim L$ and $l > L$, are approximate solutions for low momentum corresponding to large distances.  The approximate solutions are based on the trial function of the form given by eq.~(\ref{tfunc}), $\epsilon~(a) = 0$ for small~(large) $l$ solutions respectively.  Analytically, we would expect to be able to solve the gap equation, eq.~(\ref{pgap}), more precisely by using perturbative method order by order in the power series of $\lambda$.  The investigation in this section reveals that even though the exact solutions are not in the form given by the trial function we use here~(due to the mismatch of high momentum behaviour), they can be approximated quite excellently by these trial functions for large distances where chiral symmetry breaking phenomenon is relevant.  The existence of approximate solutions with large $l$ for arbitrary weak coupling $\lambda$ demonstrates that chiral symmetry breaking is generic in this nonlocal NJL model at least at the one-loop level.

\section{Effective potential at two-loop level}

\begin{figure}
\begin{center}
\includegraphics[width=6.5cm,height=5cm,angle=0]{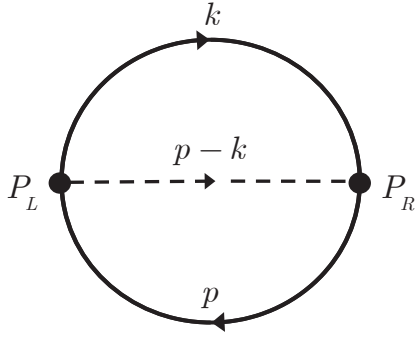}
\end{center}
\caption{Two-loop vacuum diagram for $V_{\rm 2-loop}$, solid line is the fermion and dash line is the scalar.}\label{figure2}
\end{figure}

Even though the one-loop potential demonstrates the possibility of chiral symmetry breaking solution, higher loops contribution could likewise be significant.  In this section, we will calculate the effective potential at two-loop level by following Jackiw's functional effective action method \cite{Jackiw:1974cv}. The two-loop contribution can be calculated from the vacuum expectation value of the interaction Lagrangian $\mathscr{L}_I(x)$
\begin{eqnarray}
V_{2-\rm loop} = i\,\langle\,0\,|\,\mathcal{T}\,e^{i\int {d}^4x\,\mathscr{L}_I(x)}\,|\,0\,\rangle,
\end{eqnarray}
where $\mathcal{T}$ is the time-ordering operator.  In order to obtain the two-loop contribution, we simply use
the conventional Feynman rules to calculate all possible two-loop diagrams exist in the effective theory of fermion and auxiliary scalar.  The propagator $\mathcal{G}$ of the field $\phi$ to be used in the evaluation of the 2-loop diagrams is defined by the inverse of the functional operator, namely
\begin{eqnarray}
i\,\mathcal{G}^{-1}(x,y) &=& \frac{\delta^2 }{\delta\phi(x)\,\delta\phi(y)}\int {d}^4x\,\mathscr{L}_{\rm eff}(x).
\end{eqnarray}
In the previous section, we recall the effective Lagrangian
\begin{eqnarray}
\mathscr{S}_{\rm eff} &=& \int {d}^4x\,\Bigg[i\,\bar q(x)\,\partial\!\!\!/\,q(x) + \bar q(x)\,T(x)\,P_L\,q(x) + \bar q(x)\,\bar T(x)\,P_R\,q(x)
- \frac{N_c}{\lambda}\,\frac{T(x)\,\bar T(x)}{G\left(x,\,L\right)}\Bigg],
\nonumber\\
\mathscr{S}_I &=& \int {d}^4x\,\mathscr{L}_I(x) = \int {d}^4x\,\Big[\,\bar q(x)\,T(x)\,P_L\,q(x) + \bar q(x)\,\bar T(x)\,P_R\,q(x)\,\Big].
\label{action-chiral}
\end{eqnarray}
The functional operator $\mathcal{S}^{-1}$ of the quark fields and $\mathcal{D}^{-1}$ of the complex scalar fields from the effective Lagrangian therefore can be written as
\begin{eqnarray}
i\,\mathcal{S}^{-1}(k) &=& \int {d}^4x\,e^{-\,i\,k\cdot (x-y)}\,\frac{\delta^2 \mathscr{S}_{\rm eff}}{\delta q(x)\,\delta\bar q(y)},
\nonumber\\
&=& k\!\!\!/ + T(k)\,P_L + \bar T(k)\,P_R\,,
\nonumber\\
i\,\mathcal{D}^{-1}(k) &=& \int {d}^4x\,e^{-\,i\,k\cdot (x-y)}\,\frac{\delta^2 \mathscr{S}_{\rm eff}}{\delta T(x)\,\delta\bar T(y)},
\nonumber\\
&=& -\,\frac{N_c}{\lambda\, G\left(k,\,L\right)}.  \label{sprop}
\end{eqnarray}
Noting that, $\mathcal{D}^{-1}$ has no kinetic term for the scalars $T, \bar T$\,.  The vertices of the interaction in the effective action are given by \cite{Jackiw:1974cv,Kleinert:book}
\begin{eqnarray}
\Gamma_L &=& \int {d}^4x\,e^{-\,i\,k\cdot (x-y-z)}\,\frac{(-i)^{3}\delta^3 \mathscr{S}_{\rm eff}}{\delta q(x)\,\delta\bar q(y)\,T(z)},
\nonumber\\
&=& i P_L,
\\
\Gamma_R &=& \int {d}^4x\,e^{-\,i\,k\cdot (x-y-z)}\,\frac{(-i)^{3}\delta^3 \mathscr{S}_{\rm eff}}{\delta q(x)\,\delta\bar q(y)\,\bar T(z)},
\nonumber\\
&=& i P_R\,.
\end{eqnarray}
Putting everything together, the two-loop contribution is (see figure~\ref{figure2} for the corresponding diagram)
\begin{eqnarray}
V_{2-\rm loop} &=& -\,i\,{\rm Tr}\int \frac{{d}^4k}{(2\,\pi)^4}\,\frac{{d}^4p}{(2\,\pi)^4}\,\Gamma_L\,\mathcal{S}(p)\,\Gamma_R\,\mathcal{S}(k)\,\mathcal{D}(p-k),
\nonumber\\
&=& -\frac{\lambda}{N_c}\,N_f\,N_c\int \frac{{d}^4k}{(2\,\pi)^4}\,\frac{{d}^4p}{(2\,\pi)^4}\,\frac{2\,p\cdot k\;G\left(p-k,\,L\right)}{\big[p^2 - T(p)\,\bar T(p)\,\big]
\,\big[k^2 - T(k)\,\bar T(k)\,\big]},
\end{eqnarray}
where we have used
\begin{eqnarray}
&\;&\frac{k\!\!\!/ - \bar T\,P_L - T\,P_R}{\Big(k\!\!\!/ + T\,P_L + \bar T\,P_R\Big)\,\Big(k\!\!\!/ - \bar T\,P_L - T\,P_R\Big)}
=\,  \frac{k\!\!\!/ - \bar T\,P_L - T\,P_R}{k^2 - \bar T\,T} ,
\end{eqnarray} and
\begin{eqnarray}
&&{\rm Tr}\,\Big\{P_L\,(\, p\!\!\!/ - \bar T\,P_L - T\,P_R )\,P_R\,(\,k\!\!\!/ - \bar T\,P_L - T\,P_R)\,\Big\}
=\;2\,p\cdot k.
\end{eqnarray}
Using Wick rotation i.e. $k \rightarrow i\,k_E\,,\, {d}^4k \rightarrow i\,{d}^4k_E$\,, we obtain the two-loop contribution in Euclidean space
\begin{eqnarray}
V_{\rm 2-loop}&=& -4\pi^{2}\lambda\,N_f\int \frac{{d}^4k_E}{(2\,\pi)^4}\frac{1}{\Big[\,k_E^2 + T(k_E)\,\bar T(k_E)\,\Big]}
\nonumber\\
&\;&\qquad\times\int\frac{{d}^4p_E}{(2\,\pi)^4}
\,\frac{2\,p_E\cdot k_E\,e^{-\,L\,\big|\,p_E-k_E\,\big|}}{\Big[p_E^2 + T(p_E)\,\bar T(p_E)\,\Big]\,\Big|\,p_E-k_E\,\Big|}\,.
\end{eqnarray}

The angle integration can be evaluated as shown in appendix~\ref{app2loop} to be
\begin{eqnarray}
V_{\rm 2-loop}&=& -\frac{\lambda\,N_f}{4\,L^{5}\pi^3}\int_0^{L\,\Lambda} {d}\tilde k_E\,\frac{\tilde k_E^4}{\Big[\,\tilde k_E^2 + L^2\,T(k_E)\,\bar T(k_E)\,\Big]}
\int_0^{L\,\Lambda}{d}\tilde p_E\,\frac{\tilde p_E^4}{\Big[\tilde p_E^2 + L^2\,T(p_E)\,\bar T(p_E)\,\Big]}
\nonumber\\
&\;&\times\,\sum_{n=0}^\infty\,\frac{(-\,1)^n}{n\,!}\,\Bigg\{ -\frac{\pi\,(n-1)\,A^{\frac{n-3}{2}}}{(n+3)\,(n+5)\,B}
\nonumber\\
&\;&\qquad\times
\Bigg[2\,(A^2-B^2)\; _2F_1\Big(\,\frac{3-n}{4},\frac{5-n}{4};1;\frac{B^2}{A^2}\,\Big)
\nonumber\\
&\;&\qquad\qquad +\left(B^2\,(n+2)-2\,A^2\right) \; _2F_1\Big(\,\frac{3-n}{4},\frac{5-n}{4};2;\frac{B^2}{A^2}\,\Big)\Bigg]\,\Bigg\},
\label{V2e}
\end{eqnarray}
where $\tilde k_E \equiv L\,k_E\,,\,\tilde p_E \equiv L~p_E\,,\, A\equiv \tilde p_E^2 + \tilde k_E^2\;,\, B\equiv 2\,\tilde p_E\,\tilde k_E\,.$  Henceforth, for convenience we will simply write the Euclidean momentum without a subscript.
%%%%%%%%%%%%%%%%%%%%%%%%%%%%%%%%%%%%%%%%%%%%%%%%%%%%%%%%%%%%%%%%%%%%%%%%%%%%%%%%%%%%%%%%%%%%%%%%%%
%%%%%%%%%%%%%%%%%%%%%%%%%%%%%%%%%%%%%%%%%%%%%%%%%%%%%%%%%%%%%%%%%%%%%%%%%%%%%%%%%%%%%%%%%%%%%%%%%%

\section{Results} \label{disc}

Adding all of the one-loop and two-loop contributions, the total effective potential becomes
\begin{eqnarray}
V_{\rm eff}&=& V_{\rm 1-loop} + V_{\rm 2-loop}, \nonumber \\
&=& N_{c} \int\,\frac{d^4k_E}{(2\,\pi)^4}\, \Bigg[ \frac{T(k)\bar{T}(k)}{k^{2}+T(k)\bar{T}(k)} - \ln
\left(1 + \frac{T(k)\,\bar T(k)}{k^2}\right)\Bigg] + V_{\rm 2-loop}.
\end{eqnarray}
The resulting effective potential of the scalar shows the possibility of chiral symmetry breaking at nontrivial $T(k)\neq 0$ since the sign of the one-loop contribution is opposite to the classical action of the scalar.  The 2-loop effect as given in the form of eq.~(\ref{V2e}) could be either positive or negative depending on the relative sizes of each $n$-term.  A closer investigation reveals that the $n=0$ term is the largest and it is negative.  The odd-$n$ terms are positive with smaller values than the preceding even-$n$ terms.  Consequently, the entire two-loop potential is negative.  Since the chiral symmetric solution $T=\bar{T}=0$ gives larger negative two-loop contribution than the chiral broken case~(with smaller denominator of the integrand in eq.~(\ref{V2e})).  It is thus possible that the difference of 2-loop contributions would compensate the one-loop effect and alter the true vacuum of the theory in a significant way.  We will demonstrate that the two-loop contribution is small comparing to the leading one-loop and the chiral symmetry breaking persists.  In evaluation of the momentum integrals, we will apply a UV-cutoff $\Lambda$ required in non-renormalizable effective field theory.  The cutoff will be taken to be larger than $T_0$ and smaller than $1/L$.

\subsection{1-loop}

Since both one-loop and two-loop contributions scale with the number of flavour $N_{f}$, we will simply suppress the $N_{f}$ factor henceforth.  First, we will consider 1-loop contribution of the scalar to the effective potential and demonstrate that the potential has nontrivial minima when using the ansatz solution of the gap equation as given in eq.~(\ref{Gfun}).  The 1-loop integrations, eq.~(\ref{Vone}), can be performed in two separate momentum regions and rewritten as the following
\begin{eqnarray}
\frac{V_{\rm eff}}{N_{c}} &=&  \int\,\frac{d^4k}{(2\,\pi)^4}\, \Bigg[ \frac{T(k)\bar{T}(k)}{k^{2}+T(k)\bar{T}(k)} - \ln
\left(1 + \frac{T(k)\,\bar T(k)}{k^2}\right)\Bigg], \nonumber \\
&=& \left(\int_{0}^{T_{0}} + \int_{T_{0}}^{\Lambda}\right)~\frac{d^4k}{(2\,\pi)^4}\, \Bigg[ \frac{T(k)\bar{T}(k)}{k^{2}+T(k)\bar{T}(k)} - \ln
\left(1 + \frac{T(k)\,\bar T(k)}{k^2}\right)\Bigg], \nonumber \\
&=& -\frac{T_{0}^{4}}{16\pi^{2}}\left(\ln{2}-\frac{1}{2}+2\sum_{n=1}^{\infty}\frac{(-1)^{n-1}(1-n)}{n}F_{4n-3}(LT_{0},L\Lambda;n) \right).  \label{V2}
\end{eqnarray}
where we define the function
\begin{eqnarray}
F_{4n-3}(LT_{0},L\Lambda;n)&\equiv & E_{4n-3}(2nLT_{0})-E_{4n-3}(2nL\Lambda)\left(\frac{T_{0}}{\Lambda}\right)^{4n-4}, \nonumber \\
E_{m}(z)&\equiv & \int_{1}^{\infty}\frac{e^{-zt}}{t^{m}}~dt.  \nonumber
\end{eqnarray}
The function $F_{4n-3}(LT_{0},L\Lambda;n)$ decreases very rapidly with $n$, therefore the sum in the one-loop potential, eq.~(\ref{V2}), can be approximated by truncating at finite $n$ with a high precision.

\begin{figure}[h]
        \centering
        \subfigure[]{\includegraphics[width=0.45\textwidth]{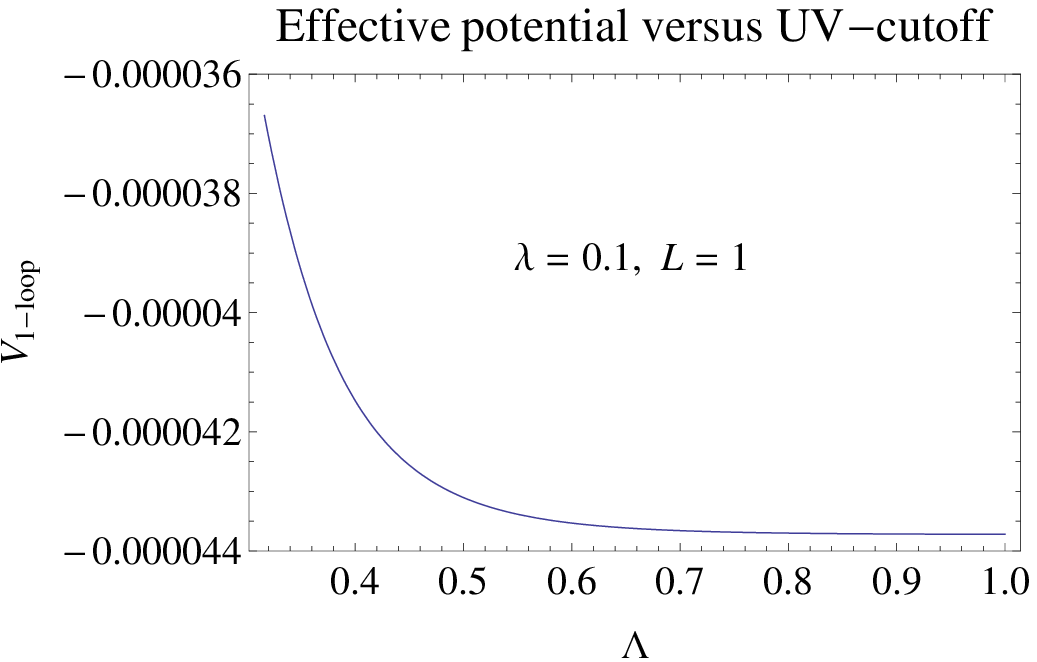}}\hfill
        \subfigure[]{\includegraphics[width=0.45\textwidth]{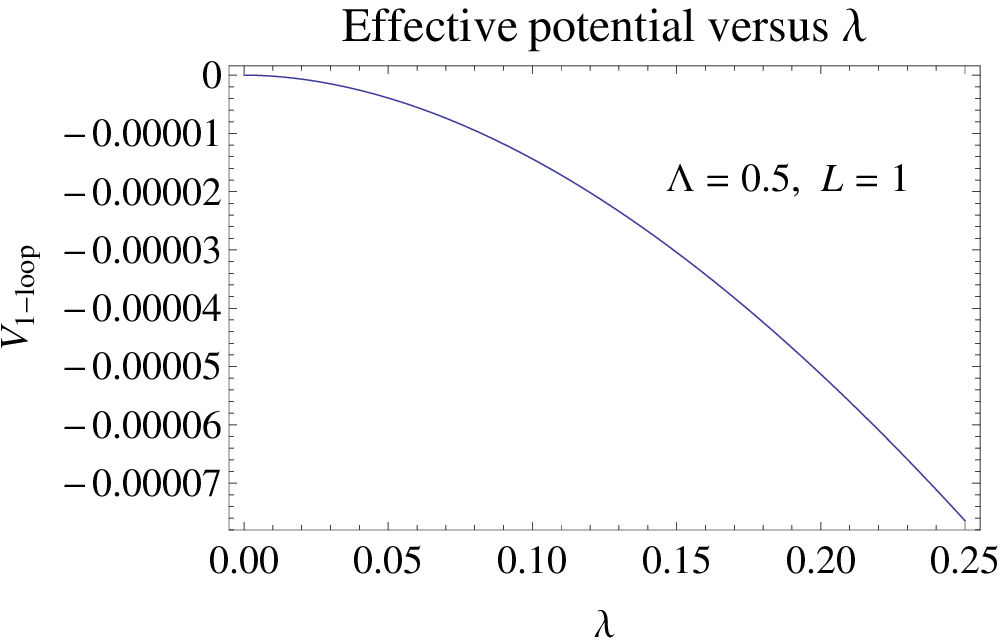}}\\
        \subfigure[]{\includegraphics[width=0.45\textwidth]{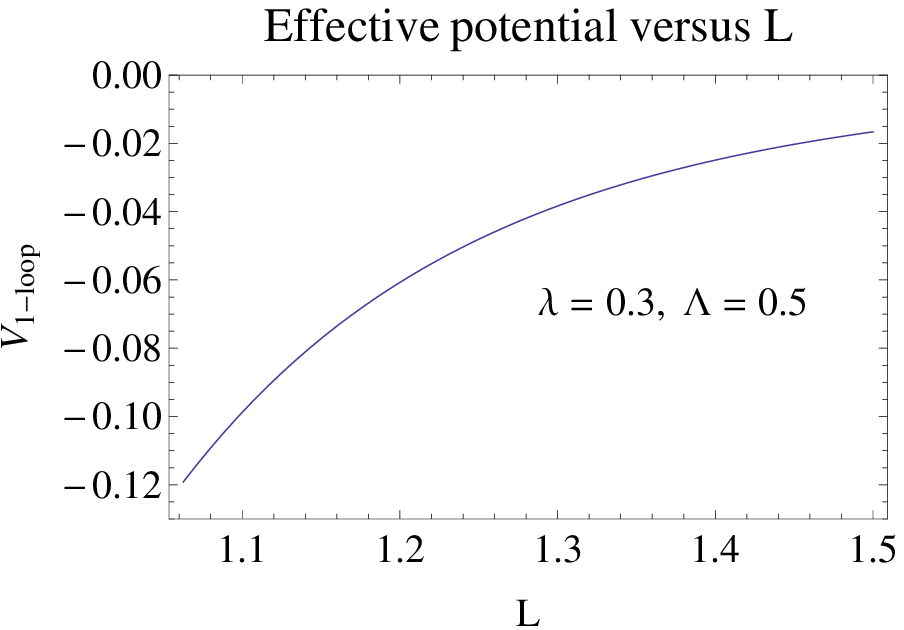}}
        \caption{One-loop effective potential per colour as a function of $\Lambda, \lambda, L$.} \label{V1fig}
\end{figure}

The one-loop contribution will be explored by fixing one and two of the 3 parameters, $\lambda, L, \Lambda$ and numerically plot the effective potential with respect to the remaining parameters.  The physically-valid region of the parameter space for our SS NJL model is $T_{0}=\sqrt{\lambda/L^{3}}<\Lambda < 1/L$.  As shown in figure~\ref{V1fig}, the one-loop potential is negative at any nonzero values of $\Lambda, \lambda, L$ corresponding to nonzero values of $T_{0}=\sqrt{\lambda/L^{3}}$.  Since when $T=0$, the potential is zero and less preferred than negative potential occuring at any coupling $\lambda$, chiral symmetry breaking thus naturally occurs for any weak coupling~(i.e. $T_{0}<1/L$).  In figure~\ref{V1fig}(a), the potential approaches negative constant for $\Lambda \gtrsim 0.7-0.8$ for $\lambda = 0.1, L=1$.  If we instead fix the UV-cutoff scale $\Lambda = 0.5$, the potential will be a decreasing function with $\lambda$ as demonstrated in figure~\ref{V1fig}(b).  Figure~\ref{V1fig}(c) also shows the one-loop potential at fixed $\lambda = 0.3, \Lambda = 0.5$ as a function of $L$.  It is important that we restrict ourselves to the physical region $T_{0}<\Lambda < 1/L$ in our consideration of the effective potential in the nonlocal NJL model.

\begin{figure}[h]
        \centering
        {\includegraphics[width=0.55\textwidth]{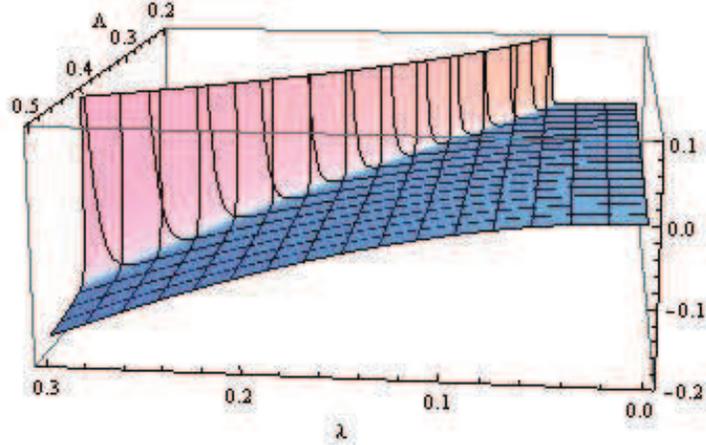}}
        \caption{One-loop effective potential per colour as a function of $\Lambda, \lambda$ for $L=1$.} \label{V1fig1}
\end{figure}

We also plot the potential landscape in the physical region at fixed $L=1$, as is shown in figure~\ref{V1fig1}.

\subsection{2-loop}
In this section, we investigate the 2-loop contribution to the effective potential.  Using the one-loop approximate scalar ansatz, eq.~(\ref{Gfun}), the two loop integration given by eq.~(\ref{V2e}) can be separated into 3 terms,
\begin{eqnarray}
V_{\rm 2-loop}& = & \frac{\lambda}{4 \pi^{3}L^{5}}\left[ \int_{0}^{T_{0}L}d\tilde{k}~\int_{0}^{T_{0}L}d\tilde{p}+ \int_{T_{0}L}^{\Lambda L}d\tilde{k}~ \int_{T_{0}L}^{\Lambda L}d\tilde{p} + 2 \int_{0}^{T_{0}L}d\tilde{k} \int_{T_{0}L}^{\Lambda L}d\tilde{p} \right] \nonumber \\
&\;&\times\,\frac{\tilde k^4}{\Big[\,\tilde k^2 + L^2\,T(\tilde{k}/L)\,\bar T(\tilde{k}/L)\,\Big]}
\frac{\tilde p^4}{\Big[\tilde p^2 + L^2\,T(\tilde{p}/L)\,\bar T(\tilde{p}/L)\,\Big]}
\nonumber\\
&\;&\times\,\sum_{n=0}^\infty\,\frac{(-\,1)^{n}}{n\,!}\,\Bigg\{ \frac{\pi\,(n-1)\,A^{\frac{n-3}{2}}}{(n+3)\,(n+5)\,B}
\nonumber\\
&\;&\qquad\times
\Bigg[2\,(A^2-B^2)\; _2F_1\Big(\,\frac{3-n}{4},\frac{5-n}{4};1;\frac{B^2}{A^2}\,\Big)
\nonumber\\
&\;&\qquad\qquad +\left(B^2\,(n+2)-2\,A^2\right) \; _2F_1\Big(\,\frac{3-n}{4},\frac{5-n}{4};2;\frac{B^2}{A^2}\,\Big)\Bigg]\,\Bigg\}.
\label{V2n}
\end{eqnarray}
The overall 2-loop contribution scales with $\lambda/L^{5}$.  The integration in the low momentum region has additional $(T_{0}L)^{6}\times (T_{0}L)^{n-1}$ factor for each $n$-term in the sum.  The integration in the high momentum region, on the other hand, has additional $(\Lambda L)^{6}\times (\Lambda L)^{n-1}$ dependence for each $n$-term.  The cross term integration has additional overall scaling factor $(T_{0}L)^{3}(\Lambda L)^{3}$ for all $n$.  We perform numerical integration on each $n$-term and add them up.  Since the integrand for each $n$ is a smooth and well-behave function with no singularities and abrupt changes, numerical integration yield very precise results.  The value of the integration for each $n$-term decreases rapidly with $n$ and the error is less than $10^{-6}$ if we truncate the sum at $n=10$.

\begin{figure}[h]
        \centering
        \subfigure[]{\includegraphics[width=0.45\textwidth]{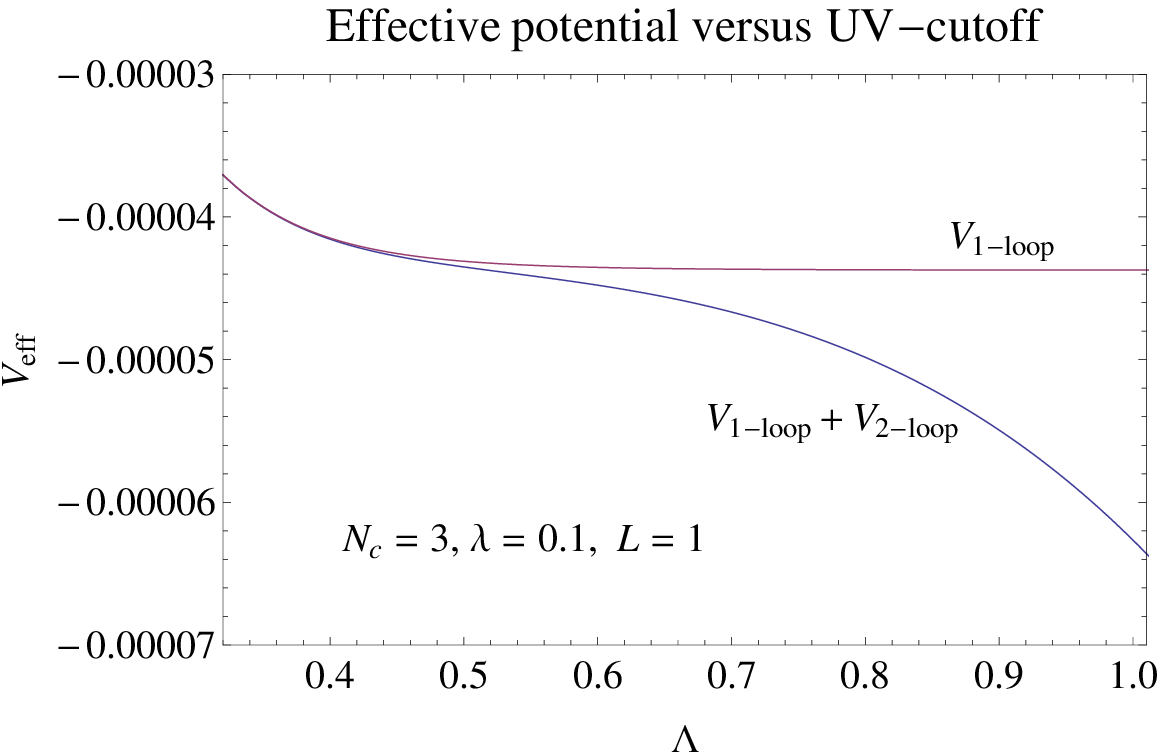}}\hfill
        \subfigure[]{\includegraphics[width=0.45\textwidth]{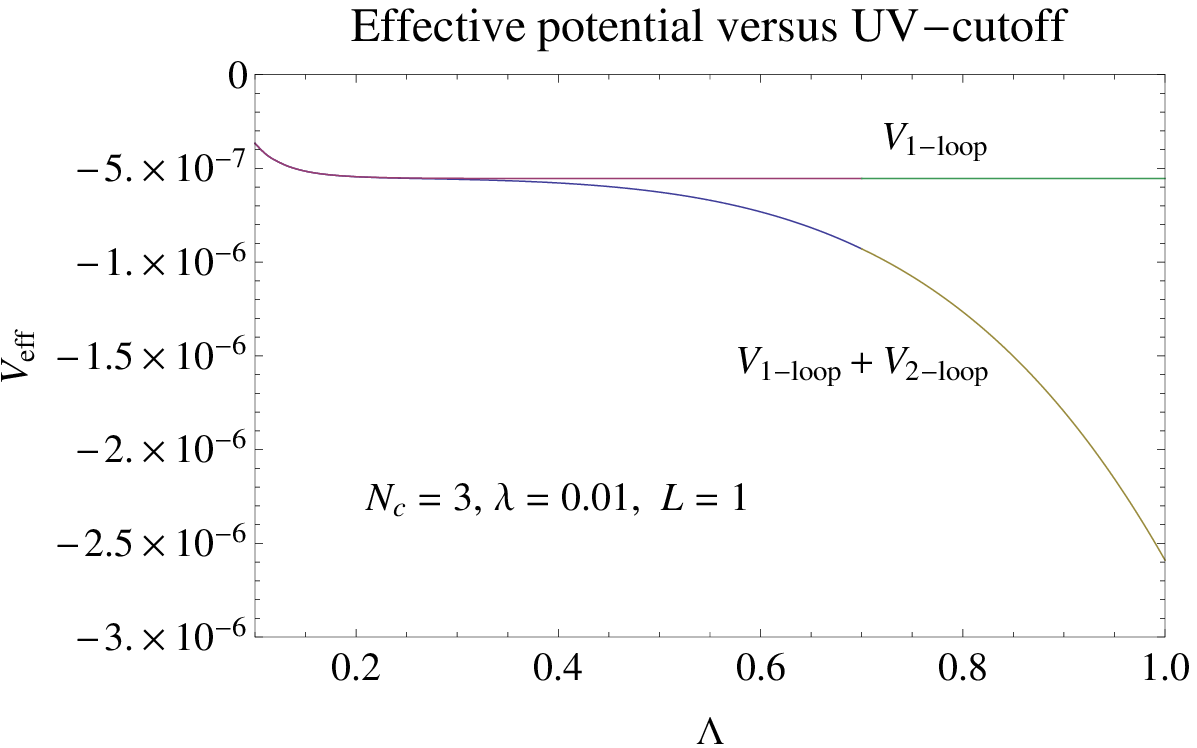}}\\
        \subfigure[]{\includegraphics[width=0.5\textwidth]{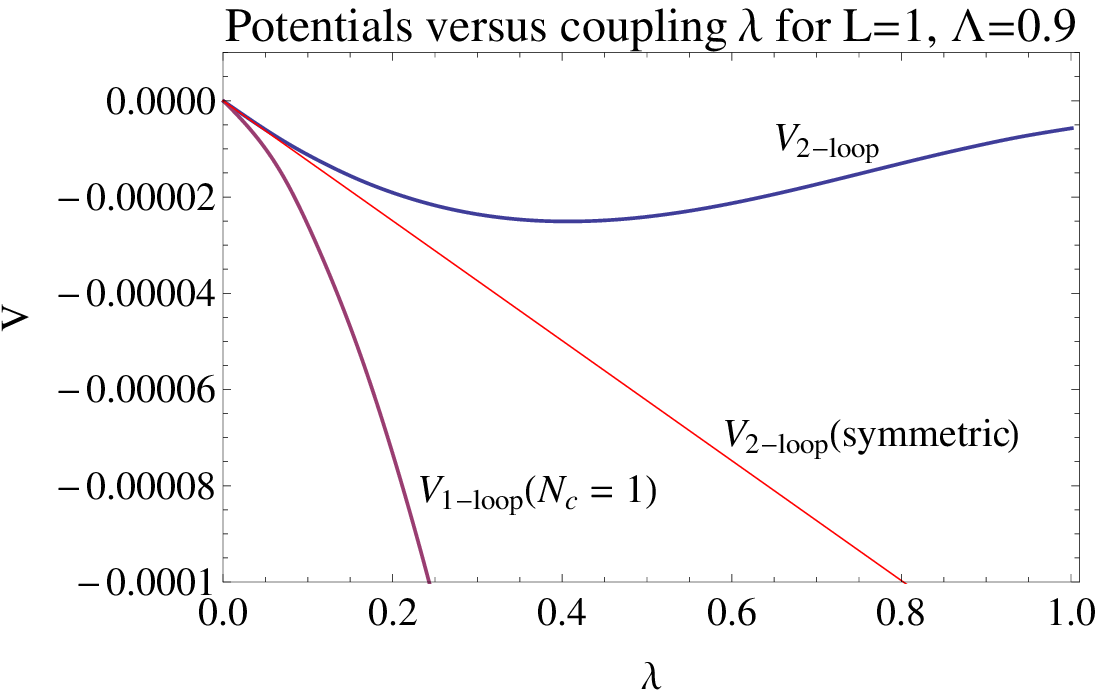}}
       \caption{Effective potential up to 2-loop as a function of $\Lambda$ for $N_{c}=3, \lambda = 0.1$~(a), $0.01$~(b); $L = 1$.  The 2-loop contributions for both chiral broken and symmetric solutions are shown in (c) in comparison to the chiral broken 1-loop. } \label{V2fig}
\end{figure}

Figure~\ref{V2fig} shows the effect of 2-loop contribution to the effective potential.  Similar to the one-loop case, the chiral-symmetry broken vacuum solution has lower energy than the chiral symmetric one~($T=\bar{T}=0$).  However, there is one crucial difference between one and two-loop potential.  The magnitude of 2-loop contribution could increase with the cutoff in contrast to the 1-loop which saturates to negative constant.  This is originated from the $\Lambda L$ dependence of the potential in eq.~(\ref{V2n}) getting larger with increasing $\Lambda$ resulting in the decreasing function of the effective potential with the cutoff.

\begin{figure}[h]
        \centering
        \subfigure[]{\includegraphics[width=0.45\textwidth]{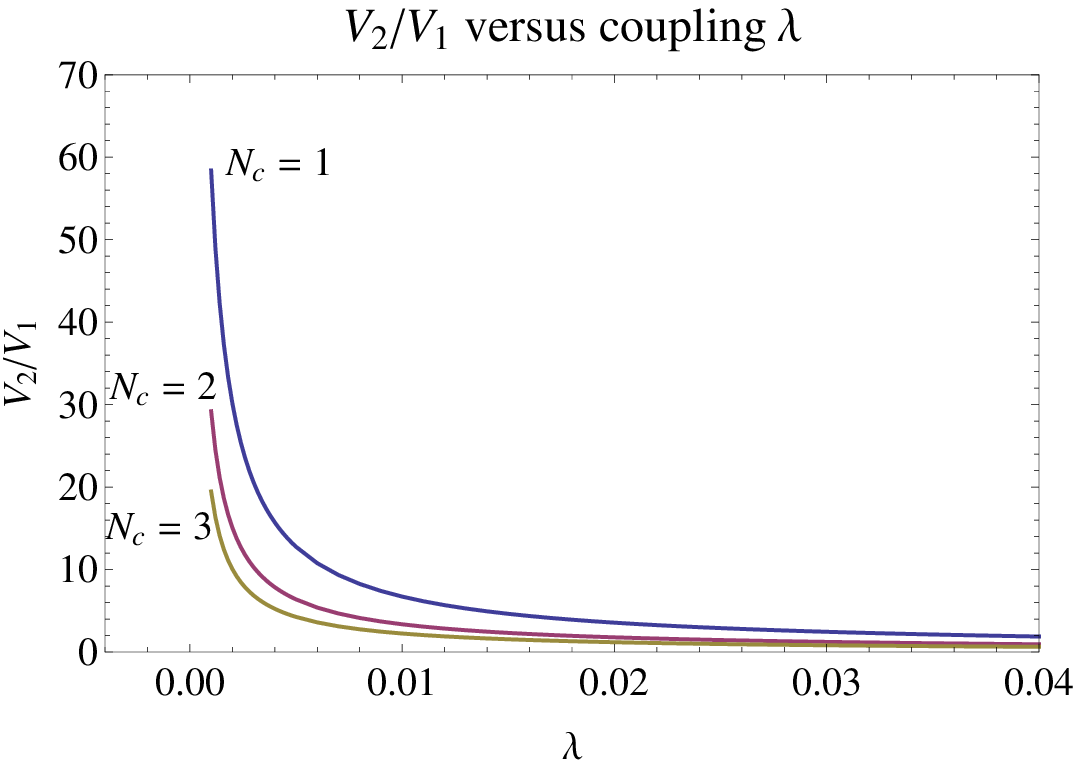}}\hfill
        \subfigure[]{\includegraphics[width=0.5\textwidth]{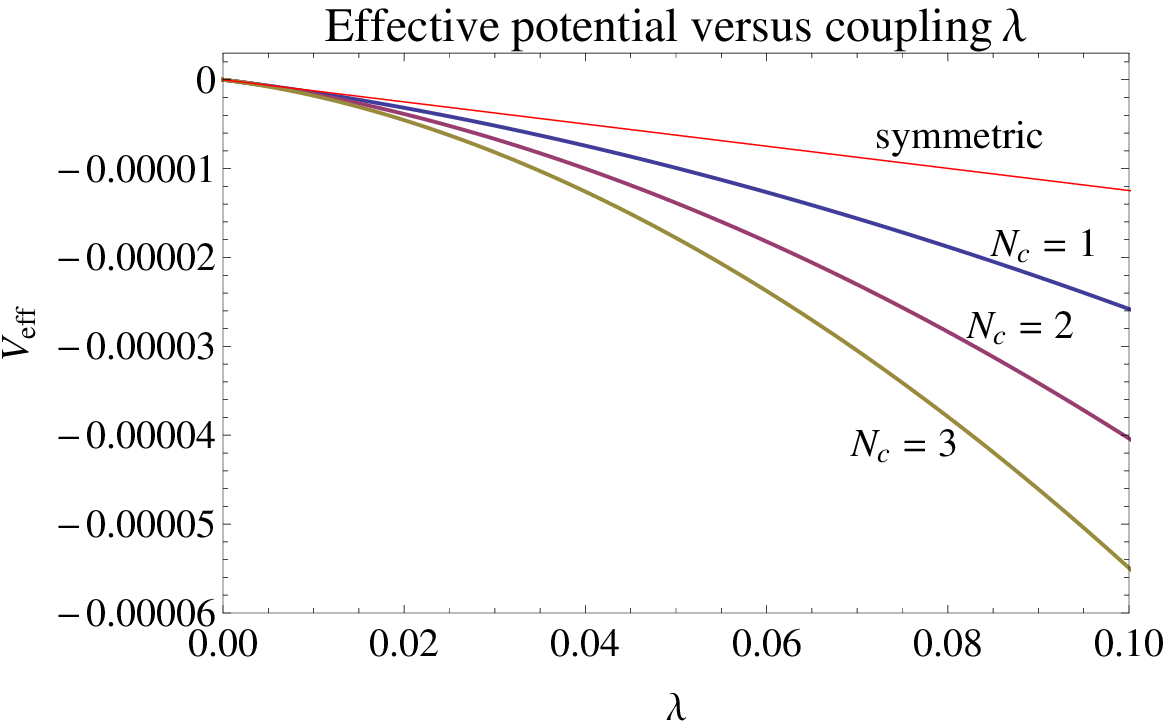}}
       \caption{The ratio of $V_{\rm 2-loop}/V_{\rm 1-loop}$ decreases rapidly with increasing $\lambda$~(a).  In (b), the effective potential for small $\lambda$ remains negative~(and smaller than the chiral symmetric case), chiral symmetry breaking thus persists. The cutoff $\Lambda$ is set to 0.9 and $L=1$ for these plots.} \label{V2fig1}
\end{figure}

When the 't Hooft coupling is very small, $\lambda \lesssim 0.08$, interesting phenomenon occurs.  As we can see from figure~\ref{V2fig1}~(a), the ratio~(magnitude) of the 2-loop to 1-loop increases sharply as $\lambda \to 0$.  From eq.~(\ref{V2}) and (\ref{V2n}), since $V_{\rm 1-loop}\sim \lambda^{2}N_{c}$ while $V_{\rm 2-loop}\sim \lambda$, the ratio of the 2-loop to 1-loop potential will scale as $(\lambda N_{c})^{-1}$ and the 2-loop contribution will be dominant for sufficiently small $\lambda$.  However, since both one and two loop contributions are negative and together they are larger in magnitude than the two-loop potential of the chiral symmetric solution~(figure~\ref{V2fig}~(c)), the chiral broken vacuum always has lower potential as is demonstrated in Figure~\ref{V2fig}~(c) and \ref{V2fig1}~(b).  Chiral symmetry breaking therefore persists for arbitrary weak coupling.  One might anticipate the chiral symmetric potential to become more negative than the chiral broken one as $\lambda \to 0$ since $V_{\rm 2-loop}(\rm sym)<V_{\rm 2-loop}(\chi SB)$ and $V_{\rm 2-loop}/V_{\rm 1-loop}\sim (N_{c}\lambda)^{-1}$.  However, a closer investigation reveals that the difference $V_{\rm 2-loop}(\rm sym)-V_{\rm 2-loop}(\chi SB)\simeq \lambda V_{\rm 2-loop}(\rm sym)$ for very small $\lambda$ and $V_{\rm 1-loop}(\chi SB)$ is actually larger in magnitude~(i.e. more negative) than $\lambda V_{\rm 2-loop}(\rm sym)$.  Consequently, the chiral broken solution still has lower potential than the chiral symmetric one even for extremely small coupling.

\section{Discussions and Conclusions}  \label{conc}

SS intersecting-branes model provides a geometrized model of chiral symmetry breaking and confinement in both weak and strong coupling regimes.  The effective field theory at low energy~($E<1/L$) from the SS model is a type of NJL model with non-local 4-fermion interaction.  In constrast to conventional NJL with 4-fermion contact interaction which requires sufficiently large coupling to break chiral symmetry, the holographic non-local NJL model prefers chiral broken phase for arbitrarily weak coupling at the 1-loop level.  In this work, we found that the 2-loop effect does NOT change this feature of the model and the chiral symmetry breaking persists for arbitrary weak coupling.  The 2-loop effect can be understood as the antiscreening of the non-local 4-fermion interaction induced from the coupling of fermion with the cloud of scalar condensate.  It is suppressed below the 1-loop contribution for $\lambda \gtrsim 0.08$.  One of the suppression factor is the number of colour degrees of freedom $N_{c}=3$ when the coupling $\lambda$ is fixed.  Bosonization of the fermion bilinear into a colour-singlet scalar naturally matches the loop expansion of the effective potential with the $1/N_{c}$ expansion.  The large-$N_{c}$ expansion makes it manifest that the 1-loop contribution scales as $(N_{c})^{1}$ and the 2-loop scales as $(N_{c})^{0}$.  The higher-order loops are therefore suppressed by negative power of $N_{c}$ and so on.

In our loop expansion (equivalent to $1/N_{c}$-expansion in this case), the higher order loops are $N_{c}$-suppressed, as well as $\lambda$ suppressed when the coupling $\lambda$ is weak.  The reason is the $\lambda/N_{c}$~(eq.~(\ref{sprop})) dependence of the scalar (condensate) propagator, the number of which counts the number of loops in the fermion loop diagrams up to a factor of $-1$, i.e. the number of loop $=$ the number of scalar propagator $+ 1$.  For example at 2-loop, there is one scalar propagator and therefore it scales with $\lambda^{1}$.  At 3-loop, there are two scalar propagators resulting in the $\lambda^{2}$-dependence of the 3-loop and so on.  It seems straightforward to see that higher loops are suppressed by higher powers of $\lambda$ in the weak coupling regime.  
     
However, the scaling of the momentum loop integral with $T_{0}$ in the 1-loop evaluation, as shown in eq.~(\ref{V2}), makes things more complicated since $T_{0}^{4}$ scales as $\lambda^{2}$.  Only when the solution of the gap equation, eq.~(\ref{Gfun}), is substituted into the one-loop potential that this extra $\lambda$-dependence appears (before the scaling, 1-loop potential scales with $\lambda^{0}$).  At 2-loop, exactly the same scaling can be done once the solution of the gap equation is substituted into the potential and we would have the extra $\lambda$-dependence as well (they are different for low and high momentum regions as we can see from eq.~(\ref{Gfun})).  This is why the 2-loop analyses are crucial in order to determine whether the chiral symmetry breaking persists at the 2-loop order.

The solution to the gap equation at one loop gives fermion condensate proportional to $\sqrt{\lambda/L^{3}}$ resulting in $\lambda^{2}$-dependence of the one-loop potential while the two-loop scales as $\lambda$.  Therefore the two-loop contribution becomes dominant to the one-loop for very small coupling.  However, since the difference between the two-loop contribution of the chiral symmetric and broken solutions is found to be numerically smaller than the size of the one-loop potential of the chiral broken solution even for very small coupling, the chiral symmetry remains broken~(Fig.~\ref{V2fig1}).

At 3-loop and higher, the situation could change since we would also have extra $\lambda$-dependence once the solution to the gap equation is substituted.  With competing loop contributions, the phase structure of the chiral symmetry could become interesting for very small coupling.  Actually this is why this model is non-trivial, the AHJK solution to the gap equation depends on $\lambda^{1/2}$ in the low momentum region while negligible in the high momentum (from eqn. (20)).  We would have extra $\lambda$-dependence in the low momentum and it could alter the conclusion on the chiral symmetry breaking of the model at lower loops.  It is likely that the difference of the higher-loop contribution between the chiral broken and symmetric configuration is again smaller than the dominant lower loops and chiral symmetry breaking persists to arbitrary order in this model.  It would be interesting to find a complete proof in the future work.

\section*{Acknowledgments}

We would like to thank Yupeng Yan, Janos Polonyi and Sanjin Benic for helpful discussions.  P.B. is supported in part by the Thailand Research Fund~(TRF) and Commission on Higher Education~(CHE) under grant RMU5380048.  D.S. is supported in part by the Thailand Center of Excellence in Physics~(ThEP).

\appendix
\section{Integrating out gauge field for single intersection model} \label{appa}
According to the standard technique of path integral (see \cite{Donoghue:1992dd}, or textbooks in QFT), the generating function of the action in eq.~(\ref{with-H}) is written by
\begin{eqnarray}
&\;&\int [d\,A_M]\,\Delta_{FP}\,\exp\left\{i\,\mathscr{S}\big(A_M,q_L\big)\right\},\nonumber\\
&\;& = \int [d\,A_M]\,\Delta_{FP}\,\exp\Big\{i\,\int\,d^5x\,\frac{1}{g_5^2}\left[\frac{1}{2}\,A_M\,\Box\,A^M + \,\delta\,(x^4)\,J^M\,A_M\right] +  i \int d^4 x\,q_L^\dagger\,\bar \sigma^\mu\,i\,\partial_\mu\,q_L\Big\}, \nonumber\\
&\;& = \int [d\,A_M]\,\Delta_{FP}
\nonumber\\
&\;&\qquad\quad\times\,\exp\,\Big\{\,\frac{i}{g_5^2}\,\int\,d^5x\,\frac{1}{2}\,A_M\,\Box\,A^M
\nonumber\\
&\;&\qquad\qquad\qquad\qquad-\,\frac{i}{g_5^2}\int\,d^5x\,\,d^5y\,\delta\,(x^4)\,\delta\,(y^4)\,\frac{1}{2}\,J^M(x)\,G_{MN}(x-y\,,\,x^4-y^4)\,J^N(y) \nonumber\\
&\;&\qquad\qquad\qquad\qquad\qquad\qquad\qquad\qquad\qquad\qquad\qquad\qquad\quad\,\, +\;  i\int d^4 x\,q_L^\dagger\,\bar \sigma^\mu\,i\,\partial_\mu\,q_L\Big\}, \label{path-H}
\end{eqnarray}
where $\Delta_{FP}$ is the Faddeev-Popov's determinant.  In the Feynman-gauge, the propagator, $G_{MN}(x,x^4)$\, of the gauge field $A_M$ can be written as
\begin{eqnarray}
G_{MN}\left(x\,,\,x^4\right) & = & \frac{1}{8\,\pi^2}\,g_{MN}\,G\left(x\,,\,x^4\right), \nonumber \\
& = & \frac{1}{8\,\pi^2}\frac{g_{MN}}{\left( (x^{4})^{2} - x^{2}\right)^{3/2}}.  \label{propa}
\end{eqnarray}
Using eq.~(\ref{without-H},\ref{path-H}) in eq.~(\ref{int-out}), we obtain
\begin{eqnarray}
e^{i\,\mathscr{S}_{\rm eff}} &=& \int\,[d\,A_M]\,\Delta_{FP}\, e^{i\,\int\,d^4x\,\mathscr{L}\big(A_M(x),q_L(x)\big)}\,\Big / \int\,[d\,A_M]\,\Delta_{FP}\, e^{i\,\int\,d^4x\,\mathscr{L}\big(A_M(x),0\big)}\nonumber\\
&=& \int [d\,A_M]\,\Delta_{FP}
\nonumber\\
&\;&\qquad\times\,\exp\,\Big\{\,\frac{i}{g_5^2}\,\int\,d^5x\,\frac{1}{2}\,A_M\,\Box\,A^M
\nonumber\\
&\;&\qquad\qquad\qquad\quad-\,\frac{i}{g_5^2}\int\,d^5x\,\,d^5y\,\delta\,(x^4)\,\delta\,(y^4)\,\frac{1}{2}\,J^M(x)\,G_{MN}(x-y\,,\,x^4-y^4)\,J^N(y) \nonumber\\
&\;&\qquad\qquad\qquad\qquad\qquad\qquad\qquad\qquad\qquad\qquad\qquad\qquad +\,  i\int d^4 x\,q_L^\dagger\,\bar \sigma^\mu\,i\,\partial_\mu\,q_L\Big\}\,
\nonumber\\
&\;&\;\Big / \int\,[d\,A_M]\,\Delta_{FP}\, \exp\Big\{\frac{i}{g_5^2}\,\int\,d^5x\,\frac{1}{2}\,A_M\,\Box\,A^M\Big\},
\nonumber\\
&=& \exp\,\Big\{-\,\frac{i}{g_5^2}\int\,d^4x\,\,d^4y\,\frac{1}{16\,\pi^2}\,g_5^2\,q_L^\dagger(x)\,\bar \sigma^\mu\,q_L(x)\,G(x-y\,,\,0)\,g_5^2\,q_L^\dagger(y)\,\bar \sigma_\mu\,q_L(y)
\nonumber\\
&\;&\qquad\qquad\qquad\qquad\qquad\qquad\qquad\qquad\qquad\qquad\qquad + \,i\int d^4 x\,q_L^\dagger\,\bar \sigma^\mu\,i\,\partial_\mu\,q_L\Big\},
\nonumber\\
&=& \exp\,\Big\{-\,i\int\,d^4x\,\,d^4y\,\frac{g_5^2}{16\,\pi^2}\,G(x-y\,,\,0)\,\Big[q_L^\dagger(x)\,\bar \sigma^\mu\,q_L(y)\Big]\,\Big[q_L^\dagger(y)\,\bar \sigma_\mu\,q_L(x)\Big]
\nonumber\\
&\;&\qquad\qquad\qquad\qquad\qquad\qquad\qquad\qquad\qquad\qquad\qquad + \,i\int d^4 x\,q_L^\dagger\,\bar \sigma^\mu\,i\,\partial_\mu\,q_L\Big\},\label{res-1}
\end{eqnarray}
where we used $J^{(4)}=0$ and Fierz identity $(q_{L,1}^\dagger\,\bar \sigma^\mu\,q_{L,2})\,(q_{L,3}^\dagger\,\bar \sigma_\mu\,q_{L,4})=(q_{L,1}^\dagger\,\bar \sigma^\mu\,q_{L,4})\,(q_{L,3}^\dagger\,\bar \sigma_\mu\,q_{L,2})$\,.

\section{Fourier transform in Euclidean $5$-dimensions} \label{appb}
The coordinates of a $d$ dimensional Euclidean space are given by
\begin{eqnarray}
x_1 &=& x\,\cos \theta_1\,,\nonumber\\
x_2 &=& x\,\sin \theta_1\,\cos \theta_2\,,\nonumber\\
x_3 &=& x\,\sin \theta_1\,\sin \theta_2\,\cos \theta_3\,,\nonumber\\
x_4 &=& x\,\sin \theta_1\,\sin \theta_2\,\sin \theta_3\,\cos \theta_4\,,\nonumber\\
x_5 &=& x\,\sin \theta_1\,\sin \theta_2\,\sin \theta_3\,\sin \theta_4\,\cos \theta_5,\nonumber\\
&\;&\vdots \nonumber\\
x_n &=& x\,\sin \theta_1\,\sin \theta_2\,\sin \theta_3\,\cdots\,\sin \theta_{n-1}\,\cos \theta_n\nonumber\\
x_{n+1} &=& x\,\sin \theta_1\,\sin \theta_2\,\sin \theta_3\,\cdots\,\sin \theta_{n-1}\,\sin \theta_n,
\end{eqnarray}
where only $\theta_{n}\in [0,2\pi)$~(so that $x_{n+1}\in[-x,+x,]$) and other angles range from $0$ to $\pi$.
For Euclidean momentum in 5 dimensions, the components are
\begin{eqnarray}
k_1 &=& k\,\cos \theta_1\,,\nonumber\\
k_2 &=& k\,\sin \theta_1\,\cos \theta_2\,,\nonumber\\
k_3 &=& k\,\sin \theta_1\,\sin \theta_2\,\cos \theta_3\,,\nonumber\\
k_4 &=& k\,\sin \theta_1\,\sin \theta_2\,\sin \theta_3\,\cos \theta_4\,,\nonumber\\
k_5 &=& k\,\sin \theta_1\,\sin \theta_2\,\sin \theta_3\,\sin \theta_4\,.
\end{eqnarray}
The volume element in 5-dimension is thus
\begin{eqnarray}
d^5 k = k^4\,\sin^3 \theta_1\,\sin^2 \theta_2\,\sin \theta_3\,{d}k\,{d}\theta_1\,{d}\theta_2\,{d}\theta_3\,{d}\theta_4.
\end{eqnarray}

With this measure, the Fourier transform in 5 dimensional Euclidean space can be performed as the following
\begin{eqnarray}
F(x) &=& \int \frac{{d}^5\,k}{(2\,\pi)^5}\,\tilde F(k)\,e^{i\,k\cdot x}=\int \frac{{d}^5\,k}{(2\,\pi)^5}\,\tilde F(k)\,e^{i\,k\,x\,\cos \theta_1},
\nonumber\\
&=& \frac{1}{(2\,\pi)^5}\int {d}k\,{d}\theta_1\,{d}\theta_2\,{d}\theta_3\,{d}\theta_4
\,k^4\,\sin^3 \theta_1\,\sin^2 \theta_2\,\sin \theta_3\,\tilde F(k)\,e^{i\,k\,x\,\cos \theta_1},
\nonumber\\
&=& \frac{1}{(2\,\pi)^5}\int_0^\infty \tilde F(k)\,k^4\,{d}k\,\int_0^\pi e^{i\,k\,x\,\cos \theta_1}\,\sin^3 \theta_1\,{d}\theta_1\,\int_0^\pi \sin^2 \theta_2\,{d}\theta_2\,\int_0^\pi \sin \theta_3\,{d}\theta_3\,\int_0^{2\,\pi} {d}\theta_4,
\nonumber\\
&=& \frac{1}{4\,\pi^3}\int_0^\infty {d}k\,\tilde F(k)\,k^4\,\Big[\frac{\sin(k\,x)}{(k\,x)^3}- \frac{\cos(k\,x)}{(k\,x)^2}\Big]\,.
\end{eqnarray}
Given an explicit functional form $\tilde F(k)$, the transform can be completed.

However, in the situation where the quarks are localized at particular $x^{4}$ and the gauge fields in 5 dimensions are integrated out to obtain the effective 4-dimensional action, we will need to perform the Fourier transform of the propagator given in eq.~(\ref{propa}) under the condition that the gauge fields are propagating at a fixed distance in $x^{4}$ direction.  The Fourier integration will split into a delta function in $x^{4}$ coordinate and the Fourier transform in the Euclidean 4 dimensions.

For example, in our model, the Fourier transform becomes
\begin{eqnarray}
G(k\,, k_4) &=& \int {d}^5\,x\,G(x\,, L)\,e^{i\,k\cdot x}
= \int_{-\,\infty}^{\infty}{d}\,x_4\,e^{i\,k_4\,x_4} \int {d}^4\,x\,G(x\,, L)\,e^{-\,i\,k\cdot\,x},
\nonumber\\
&=& \delta(k_4)\int {d}{x}\,{d}\theta_1\,{d}\theta_2\,{d}\theta_3\,
\,{x}^3\,\sin^2\theta_1\,\sin\theta_2\,G(x\,, L)\,e^{i\,{k}\,{x}\,\cos \theta_1},
\nonumber\\
&=& \delta(k_4)\int_0^\infty \frac{1}{\big(L^2 + \tilde x^2\big)^{\frac{3}{2}}}\,{x}^3\,{d}{x}\,
\frac{\pi\,J_1({k}\,{x})}{{k}\,{x}}\,(2)\,(2\,\pi),
\nonumber\\
&=& \frac{4\,\pi^2}{k}\,\delta(k_4)
\int_0^\infty \frac{({k}\,{x})^2\,J_1({k}\,{x})}{\big(({k}\,L)^2 + ({k}\,{x})^2\big)^{\frac{3}{2}}}\,{d}({k}\,{x}),
\nonumber\\
&=& \delta(k_4)\,4\,\pi^2\,\frac{e^{-\,L\,k}}{k},
\end{eqnarray}
where $J_n(x)$\,, is Bessel function.  If we neglect the momentum in bulk spacetime say, $k_4 = 0$, we obtain
\begin{eqnarray}
G(k\,, 0)&=& 4\,\pi^2\,\frac{e^{-\,L\,k}}{k}.
\end{eqnarray}

\section{Gap equation at two-loop} \label{gapeq}
In this section, we will derive the gap equation at the two-loop level.  Start with the two-loop effective potential
\begin{eqnarray}
V_{\rm eff} &=&N_{c}\left[\int\,d^4x\,T(x)\,\bar T(x)\,\frac{(x^2 + L^2)^{\frac{3}{2}}}{\lambda} -\int\,\frac{d^4k}{(2\,\pi)^4}\,\ln
\left(1 + \frac{T(k)\,\bar T(k)}{k^2}\right)\right]
\nonumber\\
&-& \lambda \int \frac{{d}^4k}{(2\,\pi)^4}\int\frac{{d}^4p}{(2\,\pi)^4}\frac{G(p-k,\,L)}{\Big[\,k^2 + T(k)\,\bar T(k)\,\Big]}
\,\frac{2\,p\cdot k}{\Big[p^2 + T(p)\,\bar T(p)\,\Big]}\,.
\end{eqnarray}

The functional derivative of $V_{\rm eff}$ with respect to $\bar T(q)$ gives the gap equation
\allowdisplaybreaks[1]
\begin{eqnarray}
\frac{\delta\,V_{\rm eff}}{\delta\,\bar T(q)} &=& \frac{\delta}{\delta\,\bar T(q)}\,\Bigg\{ N_{c}\int\,d^4x\,T(x)
\int \frac{{d}^4k}{(2\,\pi)^4}\,e^{-ik\cdot x}\,\bar T(k)\,\frac{(x^2 + L^2)^{\frac{3}{2}}}{\lambda}
\nonumber\\
&\;&\qquad\qquad\quad -\,N_{c}\int\,\frac{d^4k}{(2\,\pi)^4}\,\ln
\left(1 + \frac{T(k)\,\bar T(k)}{k^2}\right)
\nonumber\\
&\;&\qquad\qquad\qquad\qquad -\,\lambda \int \frac{{d}^4k}{(2\,\pi)^4}\int\frac{{d}^4p}{(2\,\pi)^4}\frac{G(p-k,\,L)}{\Big[\,k^2 + T(k)\,\bar T(k)\,\Big]}
\,\frac{2\,p\cdot k}{\Big[p^2 + T(p)\,\bar T(p)\,\Big]}\,\Bigg\},
\nonumber\\
&=& \frac{N_{c}}{\lambda}\int\,d^4x\,T(x)\,\int \frac{{d}^4k}{(2\,\pi)^4}\,e^{-ik\cdot x}\,\delta^{(4)}(k-q)\,(x^2 + L^2)^{\frac{3}{2}}
\nonumber\\
&\;&-\,N_{c}\int\,\frac{d^4k}{(2\,\pi)^4}\,\frac{T(k)}{k^2+ T(k)\,\bar T(k)}\,\delta^{(4)}(k-q)
\nonumber\\
&\;&-\,\lambda \int \frac{{d}^4k}{(2\,\pi)^4}\int\frac{{d}^4p}{(2\,\pi)^4}\,2\,p\cdot k\,G(p-k,\,L)\,
\nonumber\\
&\;&\times\,\Bigg(-\frac{k^2\,T(p)\,\delta^{(4)} (p-q)}{\left[ k^2 + T(k)\,\bar T(k) \right]^2 \left[ p^2+T(p)\,\bar T(p)\right]^2}
-\frac{ T(k)\,\bar T(k)\,T(p)\,\delta^{(4)} (p-q)}{\left[k^2 + T(k)\,\bar T(k) \right]^2 \left[p^2+ T(p)\,\bar  T(p)\right]^2}
\nonumber\\
&\;&\quad-\,\frac{T(p)\,\bar  T(p)\,T(k)\,\delta^{(4)} (k-q)}{\left[k^2 + T(k)\,\bar T(k) \right]^2 \left[p^2+ T(p)\,\bar  T(p)\right]^2}
-\frac{p^2\,T(k)\,\delta^{(4)} (k-q)}{\left[k^2 + T(k)\,\bar T(k) \right]^2 \left[p^2+ T(p)\,\bar  T(p)\right]^2}\,\Bigg),
\nonumber\\
&=& \frac{N_{c}}{\lambda\,(2\,\pi)^4}\int\,d^4x\,e^{-iq\cdot x}\,\frac{T(x)}{G(x,L)}
-\,\frac{N_{c}}{(2\,\pi)^4}\,\frac{T(q)}{q^2+ T(q)\,\bar T(q)}
\nonumber\\
&\;&+\,2\,\frac{\lambda}{(2\,\pi)^4} \int \frac{{d}^4k}{(2\,\pi)^4}\,
\frac{2\,q\cdot k\;G(q-k,\,L)\,k^2\,T(q)}{\left[ k^2 + T(k)\,\bar T(k) \right]^2 \left[ q^2+T(q)\,\bar T(q)\right]^2}
\nonumber\\
&\;&+\,2\,\frac{\lambda}{(2\,\pi)^4} \int \frac{{d}^4k}{(2\,\pi)^4}\,
\frac{ 2\,q\cdot k\;G(q-k,\,L)\,T(k)\,\bar T(k)\,T(q)}{\left[k^2 + T(k)\,\bar T(k) \right]^2 \left[q^2+ T(q)\,\bar  T(q)\right]^2}\,,
\end{eqnarray}
where we have used the symmetric property of the propagator $G(k,L)$\, i.e. $G(k-p,L)=G(|k-p|\,,L)=G(|p-k|\,,L)$.
One then obtains the E.O.M. $(\delta\,V_{\rm eff}/\delta\,\bar{T}(q)=0)$\, as
\begin{eqnarray}
&\;&\frac{N_{c}}{\lambda}\int\,d^4x\,e^{-iq\cdot x}\,\frac{T(x)}{G(x,L)}
-\,N_{c}\,\frac{T(q)}{q^2+ T(q)\,\bar T(q)}
\nonumber\\
&\;&\quad+\,2\,\lambda \int \frac{{d}^4k}{(2\,\pi)^4}\,2\,q\cdot k\,G(q-k,\,L)\,
\frac{k^2\,T(q)}{\left[ k^2 + T(k)\,\bar T(k) \right]^2 \left[ q^2+T(q)\,\bar T(q)\right]^2}
\nonumber\\
&\;&\quad+\,2\,\lambda \int \frac{{d}^4k}{(2\,\pi)^4}\,2\,q\cdot k\,G(q-k,\,L)\,
\frac{ T(k)\,\bar T(k)\,T(q)}{\left[k^2 + T(k)\,\bar T(k) \right]^2 \left[q^2+ T(q)\,\bar  T(q)\right]^2} = 0\,. \label{geqn}
\end{eqnarray}

\subsection{The $k^2\gg T(k)\,\bar T(k)$\, approximation}
We will approximate the gap equation in two regimes.  First when $k^2\gg T(k)\,\bar T(k)$, eq.~(\ref{geqn}) becomes
\begin{eqnarray}
&\;&\frac{N_{c}}{\lambda}\int\,d^4x\,e^{-iq\cdot x}\,\frac{T(x)}{G(x,L)}
-\,N_{c}\,\frac{T(q)}{q^2}
\nonumber\\
&\;&\qquad+\,2\,\lambda \int \frac{{d}^4k}{(2\,\pi)^4}\,2\,q\cdot k\,G(q-k,\,L)\,
\frac{k^2\,T(q)}{k^4\,q^4}
\nonumber\\
&\;&\qquad+\,2\,\lambda \int \frac{{d}^4k}{(2\,\pi)^4}\,2\,q\cdot k\,G(q-k,\,L)\,
\frac{ T(k)\,\bar T(k)\,T(q)}{k^4\,q^4} = 0.
\end{eqnarray}
Multiply by $q^{4}$ and neglect the last term in the left-hand side, we obtain
\begin{eqnarray}
\frac{N_{c}\,q^{4}}{\lambda}\int\,d^4x\,e^{-iq\cdot x}\,\frac{T(x)}{G(x,L)}
-\,N_{c}\,q^{2}T(q)
+\,2\,\lambda \int \frac{{d}^4k}{(2\,\pi)^4}\,2\,q\cdot k\,G(q-k,\,L)\,
\frac{T(q)}{k^2} = 0\,.
\end{eqnarray}
The last term represents the non-local screening effect of the scalar which is $N_{c}$-suppressed comparing to the other terms.  If we neglect the screening effect and Fourier transform the rest, the one-loop gap equation is recovered,
\begin{eqnarray}
\nabla^2\left(\frac{T(x)}{G(x,L)}\right) + \lambda T(x)=0,
\end{eqnarray}
where $\nabla^{2}$ is the Euclidean Laplacian in 4 dimensions.
\subsection{The $T(k)\,\bar T(k) \gg k^2$\, approximation}
Next, we consider to the low momentum regime i.e. $T(k)\,\bar T(k) \gg k^2$\,, the E.O.M. in this limit is given by
\begin{eqnarray}
&\;&\frac{N_{c}}{\lambda}\int\,d^4x\,e^{-iq\cdot x}\,\frac{T(x)}{G(x,L)}
-\,N_{c}\,\frac{T(q)}{T(q)\,\bar T(q)}
\nonumber\\
&\;&\qquad+\,2\,\lambda \int \frac{{d}^4k}{(2\,\pi)^4}\,2\,q\cdot k\,G(q-k,\,L)\,
\frac{k^2\,T(q)}{\left[ T(k)\,\bar T(k) \right]^2 \left[ T(q)\,\bar T(q)\right]^2}
\nonumber\\
&\;&\qquad+\,2\,\lambda \int \frac{{d}^4k}{(2\,\pi)^4}\,2\,q\cdot k\,G(q-k,\,L)\,
\frac{ T(k)\,\bar T(k)\,T(q)}{\left[T(k)\,\bar T(k) \right]^2 \left[T(q)\,\bar  T(q)\right]^2} = 0.
\end{eqnarray}
Neglecting the third term in the left-hand side, the gap equation becomes
\begin{eqnarray}
\quad\frac{N_{c}}{\lambda}\int\,d^4x\,e^{-iq\cdot x}\,\frac{T(x)}{G(x,L)}
=\,N_{c}\,\frac{1}{\bar T(q)}
-\,2\,\lambda \int \frac{{d}^4k}{(2\,\pi)^4}\,
\frac{2\,q\cdot k\,G(q-k,\,L)}{T(k)\,\bar T(k)\,T(q)\,\bar T(q)}\,\frac{1}{\bar  T(q)}\,.
\end{eqnarray}
The last term on the right-hand side represents the non-local screening effect of the scalar which is $N_{c}$-suppressed.  Again, the one-loop gap equation is recovered when the screening effect is neglected.

The gap equation at one-loop level is solved in ref.~\cite{Antonyan:2006vw} as given in eq.~(\ref{Gfun}).  We use this approximate solution in the evaluation of the effective potential.

\section{Evaluation of the 2-loop angle integration} \label{app2loop}
We will integrate out the internal angle of Euclidean 4-dimension, we start with the two-loop effective potential,
%This section will use the Legendre expansion in the scalar Green function $G(p-k,L)$\, which performed by \cite{Winkler:2010,Swiatecki:1951}.

\begin{eqnarray}
V_{\rm 2-loop} &=&  -4\pi^{2}\lambda\,N_f\int \frac{{d}^4k_E}{(2\,\pi)^4}\frac{1}{\Big[\,k_E^2 + T(k_E)\,\bar T(k_E)\,\Big]}\int\frac{{d}^4p_E}{(2\,\pi)^4}
\,\frac{2\,p_E\cdot k_E}{\Big[p_E^2 + T(p_E)\,\bar T(p_E)\,\Big]}\,\frac{e^{-\,L\,\big|\,p_E-k_E\,\big|}}{\Big|\,p_E-k_E\,\Big|}
\nonumber\\
&=& -\frac{ 4\pi^{2}\lambda\,N_f}{(2\,\pi)^8}\int_0^\Lambda {d}k_E\,\frac{(2\,\pi^2)\,k_E^4}{\Big[\,k_E^2 + T(k_E)\,\bar T(k_E)\,\Big]}
\nonumber\\
&\;&\qquad\int_0^\Lambda{d}p_E\int_0^\pi {d}\theta
\,\frac{2\,(4\,\pi)\,p_E^4\,\sin^2\theta\,\cos\theta}{\Big[p_E^2 + T(p_E)\,\bar T(p_E)\,\Big]}\,
\frac{e^{-\,L\,\sqrt{p_E^2 -2\,p_E\,k_E\,\cos\theta +k_E^2}}}{\sqrt{p_E^2 -2\,p_E\,k_E\,\cos\theta +k_E^2}}
\nonumber\\
&=& -\frac{\lambda\,N_f}{4\,L^5\,\pi^3}\int_0^{L\,\Lambda} {d}(L\,k_E)\,\frac{(L\,k_E)^4}{\Big[\,(L\,k_E)^2 + L^2\,T(k_E)\,\bar T(k_E)\,\Big]}
\nonumber\\
&\;&\times\int_0^{L\,\Lambda}{d}(L\,p_E)\,\frac{(L\,p_E)^4}{\Big[(L\,p_E)^2 + L^2\,T(p_E)\,\bar T(p_E)\,\Big]}
\nonumber\\
&\;&\qquad\times\int_{-\,1}^1 {d}(\cos\theta)\;\cos\theta\,\sqrt{1-\cos^2\theta}\;
\frac{e^{-\,\sqrt{(L\,p_E)^2 -2\,(L\,p_E)\,(L\,k_E)\,\cos\theta +(L\,k_E)^2}}}{\sqrt{(L\,p_E)^2 -2\,(L\,p_E)\,(L\,k_E)\,\cos\theta + (L\,k_E)^2}}
\nonumber\\
&=& -\frac{\lambda\,N_f}{4\,L^{5}\pi^3}\int_0^{L\,\Lambda} {d}\tilde k_E\,\frac{\tilde k_E^4}{\Big[\,\tilde k_E^2 + L^2\,T(k_E)\,\bar T(k_E)\,\Big]}
\int_0^{L\,\Lambda}{d}\tilde p_E\,\frac{\tilde p_E^4}{\Big[\tilde p_E^2 + L^2\,T(p_E)\,\bar T(p_E)\,\Big]}
\nonumber\\
&\;&\qquad\times\int_{-\,1}^1 {d}x\; x\,\sqrt{1-x^2}\;
\frac{e^{-\,\sqrt{A - B\,x}}}{\sqrt{A - B\,x}}\,,
\end{eqnarray}
where $\tilde k_E \equiv L\,k_E\,,\;\tilde p_E \equiv L~p_E\,,\; A\equiv \tilde p_E^2 + \tilde k_E^2\,,\; B\equiv 2\,\tilde p_E\,\tilde k_E$\,.
\\

Expanding function $\frac{e^{u}}{u} = \frac{1}{u}\,\sum_{n=0}^\infty \frac{u^n}{n\,!}\,$ gives
\begin{eqnarray}
V_{\rm 2-loop} &=& -\frac{\lambda\,N_f}{4\,L^{5}\pi^3}\int_0^{L\,\Lambda} {d}\tilde k_E\,\frac{\tilde k_E^4}{\Big[\,\tilde k_E^2 + L^2\,T(k_E)\,\bar T(k_E)\,\Big]}
\int_0^{L\,\Lambda}{d}\tilde p_E\,\frac{\tilde p_E^4}{\Big[\tilde p_E^2 + L^2\,T(p_E)\,\bar T(p_E)\,\Big]}
\nonumber\\
&\;&\qquad\times\int_{-\,1}^1 {d}x\; x\,\sqrt{1-x^2}\;
\,\frac{1}{\sqrt{ A - B\,x}}\,\sum_{n=0}^\infty\,\frac{(-\,1)^n}{n\,!}\left(\sqrt{ A - B\,x}\right)^n,
\nonumber\\
&=& -\frac{\lambda\,N_f}{4\,L^{5}\pi^3}\int_0^{L\,\Lambda} {d}\tilde k_E\,\frac{\tilde k_E^4}{\Big[\,\tilde k_E^2 + L^2\,T(k_E)\,\bar T(k_E)\,\Big]}
\int_0^{L\,\Lambda}{d}\tilde p_E\,\frac{\tilde p_E^4}{\Big[\tilde p_E^2 + L^2\,T(p_E)\,\bar T(p_E)\,\Big]}
\nonumber\\
&\;&\qquad\times\,\sum_{n=0}^\infty\,\frac{(-\,1)^n}{n\,!}\int_{-\,1}^1 {d}x\; x\,\sqrt{1-x^2}\;
\left(A - B\,x\right)^{\frac{n-1}{2}},
\nonumber\\
&=& -\frac{\lambda\,N_f}{4\,L^{5}\pi^3}\int_0^{L\,\Lambda} {d}\tilde k_E\,\frac{\tilde k_E^4}{\Big[\,\tilde k_E^2 + L^2\,T(k_E)\,\bar T(k_E)\,\Big]}
\int_0^{L\,\Lambda}{d}\tilde p_E\,\frac{\tilde p_E^4}{\Big[\tilde p_E^2 + L^2\,T(p_E)\,\bar T(p_E)\,\Big]}
\nonumber\\
&\;&\times\,\sum_{n=0}^\infty\,\frac{(-\,1)^n}{n\,!}\,\Bigg\{ -\frac{\pi\,(n-1)\,A^{\frac{n-3}{2}}}{(n+3)\,(n+5)\,B}
\nonumber\\
&\;&\qquad\times
\Bigg[2\,(A^2-B^2)\; _2F_1\Big(\,\frac{3-n}{4},\frac{5-n}{4};1;\frac{B^2}{A^2}\,\Big)
\nonumber\\
&\;&\qquad\qquad+\left(B^2\,(n+2)-2\,A^2\right) \; _2F_1\Big(\,\frac{3-n}{4},\frac{5-n}{4};2;\frac{B^2}{A^2}\,\Big)\Bigg]\,\Bigg\},
\end{eqnarray}
where  \,$_2F_1(a,b;c;z)$\, is the hypergeometric function.

\end{document}